\documentclass[11pt]{article}
\pdfoutput=1

\usepackage{jheppub}
\usepackage{amssymb, amsfonts, amsmath, amsthm, lineno}
\usepackage{enumerate}
\usepackage{mathrsfs}
\usepackage{bbold}
\usepackage{verbatim}
\usepackage[dvipsnames]{xcolor}
\usepackage{tikz, adjustbox}
\usepackage{tikz-cd}
\usetikzlibrary{arrows.meta, bending}
\usetikzlibrary{shapes.geometric, arrows}
\usepackage{hyperref}
\usepackage{todonotes}
\usepackage{mathtools}
\usepackage{booktabs} 
\usetikzlibrary{shapes.geometric, arrows}
\usepackage{hyperref}
\usepackage{slashed}
\usepackage{color}
\usepackage{empheq}
\usepackage{soul}
\usepackage{cancel}
\usepackage{placeins}
\usepackage{xargs}
\usepackage{dsfont}
\usepackage{comment}
\usepackage{dirtytalk}

\usepackage{mathabx,graphicx}

\makeatletter
\newcommand{\Vast}{\bBigg@{4.75}}
\makeatother

\newcommand{\be}{\begin{equation}}
\newcommand{\ee}{\end{equation}}
\newcommand{\bea}{\begin{eqnarray}}
\newcommand{\eea}{\end{eqnarray}}

\newcommand{\CC}{\mathcal{C}}

\newcommand{\CF}{\mathcal{F}}
\newcommand{\CG}{\mathcal{G}}
\newcommand{\CH}{\mathcal{H}}

\newcommand{\CN}{\mathcal{N}}
\newcommand{\CM}{\mathcal{M}}
\newcommand{\CO}{\mathcal{O}}

\newcommand{\CR}{\mathcal{R}}
\newcommand{\CT}{\mathcal{T}}

\newcommand{\lr}{\left (}
\newcommand{\rr}{\right )}
\newcommand{\ls}{\left [}
\newcommand{\rs}{\right ]}

\newcommand\qt\tau

\newcommand{\Tr}{\text{Tr}}

\newcommand{\p}{\partial}

\renewcommand{\tilde}[1]{\widetilde{#1}}

\renewcommand{\emph}[1]{\textit{#1}}

\makeatletter
\renewcommand{\@seccntformat}[1]{\csname the#1\endcsname.\,\,}
\makeatother
\allowdisplaybreaks

\let \savenumberline \numberline
\def \numberline#1{\savenumberline{#1.}}

\makeatletter
\def\@fpheader{\relax}
\makeatother

\usetikzlibrary{shapes, arrows, chains}
\usetikzlibrary{decorations.markings}
\usetikzlibrary{decorations.pathmorphing}
\usetikzlibrary{shapes.multipart}
\usetikzlibrary{positioning}
\tikzset{snake it/.style={decorate, decoration=snake}}

\newcommand{\dd}{\mathrm{d}}

\usepackage{mdframed}
\newmdenv[
topline=true, bottomline=true, rightline=true, leftline=true,
linewidth=1pt, linecolor=red!75, skipabove=1em, skipbelow=0em, backgroundcolor=red!10
]{rmrkred}
\newmdenv[
topline=true, bottomline=true, rightline=true, leftline=true,
linewidth=1pt, linecolor=vub!75, skipabove=1em, skipbelow=0em, backgroundcolor=vub!10
]{rmrkblue}
\newmdenv[
topline=true, bottomline=true, rightline=true, leftline=true,
linewidth=1pt, linecolor=PineGreen!75, skipabove=1em, skipbelow=0em, backgroundcolor=PineGreen!10
]{rmrkgreen}
\newmdenv[
topline=true, bottomline=true, rightline=true, leftline=true,
linewidth=1pt, linecolor=black!75, skipabove=1em, skipbelow=0em, backgroundcolor=black!10
]{rmrkblack}

\definecolor{vub}{RGB}{0,52,154}
\definecolor{vubo}{RGB}{255,102,0}
\definecolor{redd}{RGB}{255,30,30}
\definecolor{r}{RGB}{228,32,20}
\definecolor{o}{RGB}{238,69,4}
\definecolor{y}{RGB}{253,228,1}
\definecolor{g}{RGB}{108,160,0}
\definecolor{b}{RGB}{0,162,203}
\definecolor{i}{RGB}{120,42,117}
\definecolor{vred}{rgb}{0.78, 0.03, 0.08}

\title{\ \vspace{0.2cm} \\
Non-Lorentzian Supergravity from Matrix Theory}
\author[a]{Dawid Maskalaniec,}
\author[b,c]{Ziqi Yan,}
\author[d]{and Utku Zorba}
\emailAdd{dawid\_maskalaniec@berkeley.edu}
\emailAdd{ziqi.yan@nbi.ku.dk}
\emailAdd{utku.zorba@iuc.edu.tr}
\affiliation[a]{Leinweber Institute for Theoretical Physics and Department of Physics, 
\\
University of California, Berkeley, California 94720, U.S.A.\smallskip}
\affiliation[b]{Center of Gravity, The Niels Bohr Institute, Copenhagen University\\ 
Blegdamsvej 17, DK-2100 Copenhagen \O, Denmark \smallskip}
\affiliation[c]{Nordita, KTH Royal Institute of Technology and Stockholm University\\
Hannes Alfv\'{e}ns v\"{a}g 12, SE-106 91 Stockholm, Sweden \smallskip}

\affiliation[d]{Department of Engineering Sciences, Istanbul University-Cerrahpa\c{s}a\\
Avcilar 34320 Istanbul, T\"{u}rkiye \smallskip
}
\preprint{NORDITA 2026-032}
\abstract{It was recently shown that the decoupling limits leading to matrix (gauge) theories on D-branes give rise to non-Lorentzian target space geometries. Perturbatively, matrix theory describes a quantum gravity theory whose low-energy supergravity description exhibits non-Lorentzian behavior. Focusing on the D-particle case associated with the Banks-Fischler-Shenker-Susskind matrix theory, and using techniques from ambitwistor string theory, we show evidence that the dynamics of this non-Lorentzian gravity should be related to anomalies in the current algebra of the associated fundamental string worldsheet theory. At large $N$, the D-particle backreaction deforms the non-Lorentzian supergravity to the Lorentzian IIA theory, providing a holographic description of the BFSS matrix theory.
At a moderately large $N$ such that the D-particles decouple at the leading order, this non-Lorentzian supergravity maps holographically to the leading-order contribution of weakly coupled bulk gravity. This approximately non-Lorentzian regime is related to the null reduction of eleven-dimensional supergravity. Within the non-Lorentzian supergravity, non-trivial dynamics arises from the backreaction of extended brane objects that form BPS states with the D-particles. Finally, we  generalize these results to other D-brane and string soliton holographic constructions. 
}  

\begin{document}

\maketitle

\section{Introduction}

One attractive approach to quantum gravity comes from the idea that the geometric nature of general relativity might be emergent, as a low-energy approximation of a quantum mechanical system without gravity. An elegant and concrete realization of this idea is the AdS/CFT correspondence~\cite{Maldacena:1997re}, which is ultimately underlaid by the renowned principle of holography stating that the degrees of freedom in a gravitational system are captured by a quantum mechanical system in one lower dimension~\cite{tHooft:1993dmi, Susskind:1994vu}. 

Another remarkable realization of emergent gravity in a quantum mechanical system is the matrix theory approach proposed by Banks, Fischler, Shenker, and Susskind (BFSS)~\cite{Banks:1996vh}, which conjectures that the large-$N$ limit of the quantum mechanics of $N$ non-relativistic D-particles describes M-theory in eleven-dimensional asymptotically flat spacetime. This quantum system of nine $N \times N$ matrices is intimately related to the (second) quantization of the supermembranes in M-theory~\cite{deWit:1988wri} and appears to be surprisingly rich, especially at large $N$, from which emerges not only various extended objects in string theory but also black hole physics. Despite of great efforts dedicated to understanding how precisely that gravity emerges from the quantization of the BFSS matrix theory, many mysteries still remain. Built on the modern advances in holography, scattering amplitudes, and bootstrap methods, there has been a recent resurgence in the study of matrix theories~(see \emph{e.g.}~\cite{Han:2020bkb, Miller:2022fvc, Lin:2023owt, Tropper:2023fjr, Herderschee:2023pza, Herderschee:2023bnc, Komatsu:2024vnb, Hartnoll:2024csr, Cho:2024kxn, Lin:2024vvg, Komatsu:2024bop, Ciceri:2025maa, Biggs:2025qfh, Lin:2025srf, Ciceri:2025wpb}). While Monte Carlo simulations continue to thrive in their applications in matrix theory (see \emph{e.g.}~\cite{Anagnostopoulos:2007fw, Berkowitz:2016jlq}), modern approaches via quantum simulation and deep learning are also being developed~\cite{Rinaldi:2021jbg}. These lines of research may provide a stringy perspective on flat space holography. 

Modern developments in non-Lorentzian geometry also provide new insights for matrix theory~\cite{Andringa:2012uz, Harmark:2017rpg, Bergshoeff:2018yvt, Ebert:2023hba, Blair:2023noj, Gomis:2023eav, Lambert:2024uue, Lambert:2024yjk, Blair:2024aqz, Lambert:2024ncn, Harmark:2025ikv, Guijosa:2025mwh, Blair:2025prd, Blair:2025nno, Argandona:2025jhg}. The BFSS matrix theory arises from a decoupling limit that zooms in on the near-BPS modes of D-particles. This is a sub-string limit~\cite{Shenker:1995xq} under which the fundamental strings in type IIA superstring theory become non-vibrating, and the dynamics is instead captured by non-relativistic D-particles~\cite{Danielsson:1996uw, Kabat:1996cu, Douglas:1996yp}. In the perturbative regime of the BFSS matrix theory, the ten-dimensional target space geometry does not admit any metric description in the decoupling limit, and there is \emph{no} propagating graviton but only instantaneous Newton-like gravitational forces. This is the regime that is historically said to have `no geometry'~\cite{Polchinski:1999br} but now understood to acquire the non-Lorentzian geometric description with an absolute time direction~\cite{Blair:2023noj, Gomis:2023eav}. This geometry does not admit any ten-dimensional metric, and the appropriate language is the Newton-Cartan formalism in terms of vielbein fields (see~\cite{Hartong:2022lsy} for a modern review). In this sense, the BFSS matrix theory may be viewed as a D-particle worldline formalism for a ten-dimensional quantum gravity whose low-energy effective description is a non-Lorentzian supergravity. At large $N$, however, the backreaction of the D-particles becomes important, leading to a deformation that goes beyond the non-Lorentzian supergravity~\cite{Blair:2024aqz}. This backreaction creates a Lorentzian bulk geometry, on which we find the familiar weakly coupled type IIA supergravity dual of the strongly coupled BFSS matrix theory.\,\footnote{Unlike the AdS${}_5$/CFT${}_4$ correspondence, the holographic dual of the BFSS matrix theory is more complex. This is because there is \emph{no} conformal symmetry and the bulk dilaton field is dependent on the radial distance from the backreacted D-particles at the center~\cite{Itzhaki:1998dd}.} 

There has already been a series of studies of different non-Lorentzian (super)gravities~\cite{Bergshoeff:2018vfn, Gallegos:2020egk, Bergshoeff:2021bmc, Bergshoeff:2021tfn, Blair:2021waq, Bergshoeff:2023ogz, Bergshoeff:2024ipq, Bergshoeff:2025grj} motivated by non-relativistic string theory~\cite{Klebanov:2000pp, Gomis:2000bd, Danielsson:2000gi} (see~\cite{Oling:2022fft, Demulder:2023bux} for reviews of more recent developments), which can be viewed as the first-quantized description of matrix string theory~\cite{Motl:1997th, Dijkgraaf:1997vv}. Not only that intriguing new formal structures of such non-Lorentzian supergravities are discovered, but also non-trivial curved solutions are found~\cite{Lambert:2024uue, Lambert:2024yjk, Lambert:2024ncn, Blair:2024aqz, Harmark:2025ikv, Blair:2025ewa} (see~\cite{Bergshoeff2023} for a recent review). In particular, significant progress has been made for understanding non-relativistic string theory as a unitary and ultra-violet complete perturbative string theory. It is shown that the vanishing of the beta-functions in the associated sigma models gives rise to the equations of motion governing a stringy generalization of Newton-Cartan supergravity~\cite{Gomis:2019zyu, Yan:2019xsf, Gallegos:2019icg, Yan:2021lbe}. In view of that matrix string theory is dual to the BFSS matrix theory, similar duality transformations also connect general matrix (gauge) theories that arise from the decoupling limits zooming in on various D-brane configurations. On the other hand, applying these decoupling limits to supergravity leads to a zoo of (non-)Lorentzian target space geometries~\cite{Blair:2023noj, Gomis:2023eav, Blair:2024aqz}. 
Akin to the D-particle case, the backreaction of the fundamental degrees of freedom at large $N$ deforms the target space to possess a Lorentzian geometric structure~\cite{Blair:2024aqz, Harmark:2025ikv}. In the special case of matrix string theory, this deformation is closely related to the $T\bar{T}$ deformation in two dimensions~\cite{Blair:2020ops, Blair:2024aqz}. 

The presence of a sector approximated by non-Lorentzian supergravity in the BPS decoupling limit raises a series of curious questions, which we aim to address in this paper:
\begin{itemize}
    \item \emph{Is there a non-Lorentzian corner of the supergravity dual to BFSS matrix theory?
    \item \emph{What is the dynamics in non-Lorentzian supergravity?}
    \item Is there an analog of the string worldsheet formalism encoding the target space non-Lorentzian gravitational dynamics?}    
\end{itemize}
We will approach these questions by first focusing on the D-particle case, before generalizing to other D-branes and the F-string.

\vspace{3mm}

\noindent \textbf{Summary of Main Results.} In Section~\ref{sec:nlrdph} we focus on the \emph{first question} about the role played by the non-Lorentzian supergravity in holography. We start with the D-particle holography and isolate a non-Lorentzian regime in the bulk gravity dual. Truncating at the leading order, we obtain a non-Lorentzian supergravity action as in~\cite{Lambert:2024ncn} that is invariant under Galilei boosts and an emergent dilatation symmetry~\cite{Bergshoeff:2018yvt, Bergshoeff:2021bmc, Bergshoeff:2024ipq}. 
We then argue that this truncated non-Lorentzian supergravity corresponds to the regime of the BFSS matrix theory at a `moderately large' $N$. This truncation corresponds to the null reduction of eleven-dimensional supergravity. In Section~\ref{eq:nllst} we study a BPS decoupling limit of type IIA superstring theory without D-particle backreaction, which leads to a quantum gravity whose low-energy description is non-Lorentzian. In contrast, the D-particle backreaction deforms this corner to the full IIA theory. 
In connection with the \emph{third question}, we use the techniques from ambitwistor string theory~\cite{Mason:2013sva, Adamo:2014wea} to argue that the dynamics of non-Lorentzian supergravity should be related to current algebra anomalies, described by a curved $\beta\gamma$ system. In Section~\ref{sec:bth} we combine the knowledge from the previous sections to complete the discussion of holography initiated in Section~\ref{sec:nlrdph} to include the Ramond-Ramond (RR) sector, which allows us to study the brane sources in the truncated non-Lorentzian supergravity. We show that non-trivial dynamics arises from sourcing the supergravity using extended objects, and D-particles are decoupled. Finally, we generalize to other D-brane and string solution holographic constructions.
In Appendix~\ref{app:eom} we give the variations of the non-Lorentzian supergravity in the particle case. In Appendix~\ref{app:pnrst} we review the parallel studies of matrix string theory, which further strengthens our discussion of the BFSS matrix theory from the dual perspective of perturbative non-relativistic string theory. 

\section{Non-Lorentzian Regime in D-Particle Holography} \label{sec:nlrdph}

In this section we illustrate how a non-Lorentzian sector naturally arises in gauge/gravity duality by focusing on the D-particle case. After a brief review of the D-particle holography, we first consider the bulk gravity and discuss the central elements of the resulting non-Lorentzian supergravity. We will then study the correspondence of this non-Lorentzian regime in the BFSS matrix theory. 

\subsection{Review of D-Particle Holography} \label{sec:dph}

We start with reviewing the classic construction~\cite{Itzhaki:1998dd} of the correspondence between the BFSS matrix theory at large $N$ and supergravity on a background geometry containing a conformally AdS${}_2$ sector and an 8-sphere. This will help us set up the notation and, furthermore, identify a non-Lorentzian regime corresponding to the null reduction of eleven-dimensional supergravity in later subsections. 

We start with the string picture by considering a stack of $N$ coinciding D-particles. Perturbative string theory can be trusted when the string coupling $G^{}_\text{s}$ is small, and when $N$ is small such that the backreaction of the D-particles can be ignored. Denote the string length as $L^{}_\text{s}$\,. The field theory limit is
\be \label{eq:ftl}
	L^{}_\text{s} \rightarrow 0\,,
		\qquad%
	g^{2}_\text{YM} \equiv \frac{G^{}_\text{s}}{L^{3}_\text{s}} = \text{fixed}\,,
\ee
where $g^{}_\text{YM}$ characterizes the Yang-Mills (YM) coupling associated with the gauge field on the D-particles, which is dimensionful. It follows that the effective coupling of the resulting field theory is
\be \label{eq:geff}
	g^{2}_\text{eff} \sim \frac{N g^{2}_\text{YM}}{E^3}\,.
\ee
Here, $E$ is the characteristic energy scale that should be kept finite in the limit~\eqref{eq:ftl}. For example, in a physical process of the (eikonal) scattering between two bound states of D-particles, $E$ is captured by the ground-state open string exchanged between the bound-states of the D-particles separated by a characteristic length scale $R$, \emph{i.e.},
\be \label{eq:ce}
	E \sim \frac{R}{L^2_\text{s}} = \text{fixed}.
\ee 
In the field theoretic language, the finiteness of this $E$ guarantees that the Higgs expectation value is finite. Under the above limiting prescription, the ten-dimensional Newton constant goes to zero, which implies that propagating gravitational modes are decoupled. The resulting field theory is famously the $\CN = 16$ supersymmetric quantum mechanics of 9 $N \times N$ matrices~\cite{Baake:1984ie, Flume:1984mn, Claudson:1984th}, \emph{i.e.}~the BFSS matrix theory. 

It is useful to reformulate the decoupling limit using a dimensionless variable $\omega$, by first rescaling the parameters as follows:
\be \label{eq:dlo}
	L^{}_\text{s} = \omega^{-\frac{1}{2}} \, \ell^{}_\text{s}\,,
		\qquad%
	G^{}_\text{s} = \omega^{-\frac{3}{2}} \, g^{}_\text{s}\,,
		\qquad%
	R = \omega^{-1} \, r\,.
\ee
The decoupling limit defined by Eqs.~\eqref{eq:ftl} and \eqref{eq:ce} is now translated to be the infinite $\omega$ limit. Note that $R / L_\text{s} \rightarrow 0$ in this limit, \emph{i.e.}, the distance between the D-particles becomes hierarchically much smaller than the string scale. 

When $N$ is large, the effective coupling~\eqref{eq:geff} is also large and the BFSS matrix theory becomes strongly coupled. However, if the characteristic energy $E$ falls within an appropriate range, a weakly-coupled gravitational description becomes valid. To be specific, we return to the eikonal scattering between two bound states of D-particles, one consisting of a single D-particle and the other $N$ D-particles. At large $N$, the backreaction of the $N$ D-particles in the second bound state cannot be ignored. Before the decoupling limit is taken, the $N$ D-particles backreact to create the D-particle geometry,
\begin{align} \label{eq:bnhg}
	\dd s^2_{10} = - \frac{\dd t^2}{\sqrt{H}} + \sqrt{H} \, \Bigl( \dd R^2 + R^2 \, \dd \Omega_8^2 \Bigr)\,, 
        \qquad%
	e^\Phi = G^{}_\text{s} \, H^{\frac{3}{4}},
		\qquad%
	C^{(1)} = \frac{\dd t}{G^{}_\text{s} \, H},
\end{align}
with $\dd\Omega_8^2$ the line element on an 8-sphere and the harmonic function
\be
	H = 1 + \bigl(2\sqrt{\pi}\bigr)^5 \, \Gamma\bigl(\tfrac{7}{2}\bigr) \, G^{}_\text{s} \, N \, \frac{L_\text{s}^{7}}{R^{7}}\,.
\ee
Moreover, $t$ is the time direction in the target space, $R$ the radius of the 8-sphere, $C^{(1)}$ the RR one-form, and $\Phi$ the dilaton. 
Plugging the prescription~\eqref{eq:dlo} into Eq.~\eqref{eq:bnhg}, followed by sending $\omega$ to infinity, we are led to the following near-horizon geometry measured in the rescaled string length $\ell_\text{s}^{}$\,:
\begin{align} \label{eq:nhg0}
	\lr \frac{\ell_\text{s}}{L_\text{s}} \rr^{\!\!2} \dd s^2_{10} \rightarrow - \frac{\dd t^2}{\sqrt{h}} + \sqrt{h} \, \Bigl( \dd r^2 + r^2 \, \dd \Omega^2_8 \Bigr), 
        \qquad%
	e^\Phi \rightarrow g^{}_\text{s} \, h^{\frac{3}{4}},
		\qquad%
	\lr \frac{\ell_\text{s}}{L_\text{s}} \rr C^{(1)} \rightarrow \frac{\dd t}{g^{}_\text{s} \, h},
\end{align}
with the harmonic function $h$ taking the following `near-horizon' form:
\be
	h = \bigl(2\sqrt{\pi}\bigr)^5 \, \Gamma\bigl(\tfrac{7}{2}\bigr) \, g^{}_\text{s} \, N \, \frac{\ell_\text{s}^{7}}{r^{7}}\,,
		\qquad%
	r^2 = x_1^2 + \cdots + x_9^2\,. \label{eq:hfnh}
\ee
Focusing on the $(t, r)$-sector, the relevant part of the metric in Eq.~\eqref{eq:nhg0} is conformally AdS${}_2$.  
Instead of having the scattering between two D-particle bound states, from the bulk perspective, we consider a probe D-particle moving in the background geometry~\eqref{eq:nhg0}, which experiences an effective potential due to the background geometry with the characteristic length scale $r$, as the original $N$ D-particle source sits at the center of the bulk with $r = 0$~\cite{Becker:1997xw}.
The curvature $\CR$ associated with the metric~\eqref{eq:nhg0} is 
\be \label{eq:nhgc0}
	\CR \, \ell^2_\text{s} \sim \frac{1}{g^{}_\text{eff}} \sim \frac{1}{\sqrt{N \, g^{}_\text{s}}} \lr \frac{r}{\ell^{}_\text{s}} \rr^{\!\!\frac{3}{2}}\!,
\ee
which is related to the curvature radius $h^\frac{1}{4} \, r$ of the 8-sphere in Eq.~\eqref{eq:nhg0}. 
The theory is approximated by a non-stringy, weakly-coupled gravitational system if both the curvature~\eqref{eq:nhgc0} and the bulk string coupling $e^\Phi$ in Eq.~\eqref{eq:nhg0} are small, \emph{i.e.},
\be \label{eq:range0}
	\CR \, \ell^2_\text{s} \ll 1\,,
		\qquad%
	e^\Phi \ll 1
		\qquad\implies\qquad%
	N^{\frac{1}{7}}_{\phantom{s}} \ll \frac{r}{\ell^{}_{11}} \ll N^{\!\frac{1}{3}}_{\phantom{s}},
\ee
where we have introduced the Planck scale $\ell^{}_{11} \equiv g^{1/3}_\text{s} \ell^{}_\text{s}$ in M-theory. 
This condition implies,
\be
	\frac{1}{(N g^{}_\text{s})^{4/3}} \ll h \ll \frac{1}{g_\text{s}^{4/3}}\,,
\ee
For small $g^{}_\text{s}$ and large $N g^{}_\text{s}$\,, $h$ remains finite and thus the metric description~\eqref{eq:nhg0} is valid. 

In the regime $r/l_{11}\lesssim N^{1/7}$ the bulk dilaton $\Phi$ becomes large and we thus switch to M-theory on the lifted eleven-dimensional metric,
\begin{align} \label{eq:edm}
	\dd s^2_{11} &= e^{\frac{4}{3}\Phi} \bigl( g^{-2/3}_\text{s} \, \dd u + C^{}_\mu \, \dd x^\mu \bigr)^2 + e^{-\frac{2}{3}\Phi} \, \dd s^2_{10}\,,
\end{align}
with the associated (rescaled) data given in Eq.~\eqref{eq:nhg0}. Here, $u \sim u + 2\pi R_\text{s}$ is the M-theory circle. We now have to measure in terms of the eleven-dimensional Planck constant $\ell_{11}$ instead of the string length $\ell_\text{s}$\,, which requires further rescalings $(t\,,\, x^i) \rightarrow (\ell_{11} / \ell_\text{s}) (t\,,\, x^i)$\,. As a result, the metric~\eqref{eq:edm} becomes the pp-wave,
\be \label{eq:ppw}
	\dd s^2_{11} = 2 \, \dd t \, \dd u + h \, \dd u^2 + \dd r^2 + r^2 \, \dd \Omega^2_8\,.
\ee 
This pp-wave metric can be thought of as smearing in $u$ the geometry from backreacting a supergraviton in eleven dimensions. However, this smearing becomes invalid when the characteristic scale $r$ of a physical process, such as the D-particle scattering, is comparable to (or smaller than) the local radius of the M-theory circle, \emph{i.e.},
\be
	\frac{r}{\ell^{}_{11}} \lesssim e^{\frac{2}{3} \Phi} 
		\quad\implies\quad
	\frac{r}{\ell^{}_{11}} \lesssim N^{\frac{1}{9}}_{\phantom{s}}\,. 
\ee
This is because the generic sources satisfying the periodic boundary condition in $u$ are not uniformly distributed, but instead localized in $u$\,. In this regime, the BFSS matrix theory is expected to describe the scattering of supergravitons in an asymptotically flat background, providing a holographic interpretation of the BFSS conjecture. Further note that the pp-wave geometry is related to an infinite boost of a black string in the thermal case. When $r / \ell_{11} \lesssim N^{1/9}$, the black string is expected to break down into point-like particles due to the Gregory-Laflamme instability~\cite{Horowitz:1997fr}. Finally, when $r \lesssim \ell_{11}$\,, any classical metric description is expected to break down. 

\subsection{A Non-Lorentzian Regime?} \label{sec:nlr}

We now return to the metric~\eqref{eq:nhg0} containing a conformally AdS${}_2$ sector.  The holographic dual of the BFSS matrix theory as a weakly-coupled gravity theory is valid as long as the condition in Eq.~\eqref{eq:range0}
holds for the bulk curvature $\CR$ and dilaton $\Phi$\,. 
Curiously, if one further takes $\ell^{}_\text{s}$ to be small and $g^{}_\text{s}$ to be large, while keeping the eleven-dimensional Planck scale $\ell^{}_{11} = g^{1/3}_\text{s} \, \ell^{}_\text{s}$ fixed, then neither the curvature $\CR \, \ell_\text{s}^2$ \emph{nor} the bulk string coupling $e^\Phi$ changes. This implies that there is still \emph{no} extra stringy effect entering the bulk description. Under such a limit the condition~\eqref{eq:range0} remains unchanged.
To describe this limit, we introduce a new parameter $\tilde{\omega}$ via  
\be \label{eq:gslsrs0}
	g^{}_\text{s} = \tilde{\omega}^{\frac{3}{2}} \, \tilde{g}^{}_\text{s}\,,
		\qquad%
	\ell^{}_\text{s} = \frac{\tilde{\ell}^{}_\text{s}}{\tilde{\omega}^{\frac{1}{2}}}\,. 
\ee
No matter how large $\tilde{\omega}$ is, the condition~\eqref{eq:range0} continues to hold. 
Nevertheless, the large $\tilde{\omega}$ has an interesting effect on the bulk metric and RR potential. 
Note that the harmonic function scales  in $\tilde{\omega}$ as
$h = \tilde{\omega}^{-2} \, \tilde{h}$\,, with $\tilde{h} \sim N \, \tilde{g}^{}_\text{s} \, (\tilde{\ell}_\text{s}/r)^{7}$.
The near-horizon geometry~\eqref{eq:nhg0} is reparametrized as
\begin{align} \label{eq:rsnhg0}
	\dd s^2_{10} = - \frac{\tilde{\omega}}{\tilde{h}^{\frac{1}{2}}} \, \dd t^2 + \frac{\tilde{h}^{\frac{1}{2}}}{\tilde{\omega}} \, \bigl( \dd x^2_{1} + \cdots + \dd x_9^2 \bigr)\,,
        \qquad%
    e^\Phi = \tilde{g}^{}_\text{s} \, \tilde{h}^{\frac{3}{4}},
		\qquad%
	C^{(1)} \rightarrow \frac{\sqrt{\tilde{\omega}}}{\tilde{g}^{}_\text{s} \, \tilde{h}}\,\dd t\,.
\end{align}
At large $\tilde{\omega}$\,, and if 
(the D-brane number) $N$ is only `moderately large' such that it is still significantly smaller than $\tilde{\omega}^2$, \emph{i.e.},
\be
    1 \ll N \ll \tilde{\omega}^2, 
\ee
the ten-dimensional metric becomes singular. This suggests that a certain regime of the BFSS matrix theory at large $N$ has a dual bulk gravity description with non-Lorentzian features, \emph{i.e.}~without any valid ten-dimensional metric description. 

However, as long as the conditions in Eq.~\eqref{eq:range0} are satisfied, we know that the gravity dual of the BFSS matrix theory should be IIA supergravity on the background geometry~\eqref{eq:nhg0}. In what sense do we have a non-Lorentzian regime? When $\tilde{\omega}$ is large, an expansion with respect to a large $\tilde{\omega}$ can be taken in IIA supergravity. We rewrite Eq.~\eqref{eq:rsnhg0} as
\be
	\dd s^2 = \Omega \, \Bigl[ - \tilde{\omega}^2 \, \dd t^2 + \tilde{h} \, \bigl( \dd x_1^2 + \cdots + \dd x_9^2 \bigr) \Bigr]\,,
\ee
where $\Omega$ is a conformal factor. The large $\tilde{\omega}$ regime is then associated with a large speed of light regime, which implies that the leading order term in the large $\tilde{\omega}$ expansion behaves non-relativistically. 

Before proceeding further to understand the dynamics at large $\tilde{\omega}$\,, we first covariantize the $\tilde{\omega}$-reparametrization~\eqref{eq:rsnhg0} by introducing a temporal vielbein field $\tau^{}_\mu$ and transverse vielbein field $e^{}_\mu{}^i$, $i = 1\,,\,\cdots,\,9$\,, whose vacuum expectation values are 
\be \label{eq:vevte}
	\langle \tau^{}_\mu \rangle \, \dd x^\mu = \tilde{h}^{-\frac{1}{4}} \, \dd t\,,
		\qquad%
	\langle e^{}_\mu{}^i \rangle \, \dd x^\mu = \tilde{h}^{\frac{1}{4}} \, \dd x^i\,. 
\ee
We thus covariantize the prescription~\eqref{eq:rsnhg0} to be, together with the rescaling $\ell_\text{s} = \tilde{\ell}_\text{s} / \sqrt{\tilde{\omega}}$\,, 
\begin{align} \label{eq:ftoa1}
	\dd s^2 &= - \tilde{\omega} \, \bigl( \tau^{}_\mu \, \dd x^\mu \bigr)^2 + \frac{1}{\tilde{\omega}} \, e^{}_\mu{}^i \, e^{}_\nu{}^i \, \dd x^\mu \, \dd x^\nu,
        \quad\,\,%
    e^\Phi = e^\varphi,
		\quad\,\,%
	C^{(1)} = \frac{\sqrt{\tilde{\omega}}}{e^{\varphi}} \, \tau^{}_\mu \, \dd x^\mu + \frac{c^{(1)}}{{\tilde{\omega}}^{3/2}}\,,
\end{align}
with $\langle e^\varphi \rangle = \tilde{g}^{}_\text{s} \, \tilde{h}^{3/4}$. Here, $\varphi$ and $c^{(1)}$ will respectively be the dilaton and RR one-form in the non-Lorentzian supergravity that we will derive later. The $\tilde{\omega}^{-3/2}$ factor in front of $c^{(1)}$ is an educated guess at the moment, since it is expected that the subleading term is at order $O(\tilde{\omega}^{-2})$ when compared with the leading order term, just like the case for the line element. This expectation will later be corroborated in Section~\ref{sec:ddp}.

Admittedly, we are agnostic about how the Kalb-Ramond field $B^{(2)}$, RR potential $C^{(q)}$ with $q\neq1$, and open string gauge potential should be parametrized, which we will derive later in Section~\ref{sec:ddp}. Nevertheless, we do know that these reparametrizations are simply overall rescalings in $\tilde{\omega}^\alpha$, with an unknown exponent $\alpha$. Therefore, it is safe to set these extra fields to zero, which we enforce in the rest of this section.  
This allows us to study the dynamics associated with the expansion of IIA supergravity with respect to a large $\tilde{\omega}$\,, focusing on the sector consisting of the metric $G^{}_{\mu\nu}$\,, dilaton $\Phi$\,, and RR one-form $C^{(1)}$. 

\subsection{Non-Lorentzian Expansion of Supergravity} \label{sec:nlse}

We now use the ansatz~\eqref{eq:ftoa1} to expand IIA supergravity with respect to a large $\tilde{\omega}$\,. As we have just explained, we set for simplicity $B^{(2)} = C^{(q)} = 0$ with $q \neq 1$\,. Our starting point is the bosonic part of the IIA supergravity action in string frame, 
\begin{align} \label{eq:siia}
	S^{}_{\text{\scalebox{0.8}{IIA}}} &= \frac{1}{2 \, \kappa^2} \int \dd^{10} x \, \sqrt{-G} \, \biggl[ e^{-2\Phi} \Bigl( R + 4 \, \p_\mu \Phi \, \p^\mu \Phi \Bigr) -\tfrac{1}{4} \, F^{}_{\mu\nu} \, F^{\mu\nu} \biggr]\,.
\end{align}
Here, $R$ is the Ricci scalar associated with the metric $G_{\mu\nu}$ and $F_{\mu\nu} = \p_\mu C_\nu - \p_\nu C_\mu$\,. 
Note that the gravitational coupling $\kappa^2 \sim \ell^{\,8}_\text{s}$ does not contain any powers of the string coupling constant. 
Plug the ansatz~\eqref{eq:ftoa1} with $\ell_\text{s} = \tilde{\ell}_\text{s} / \sqrt{\tilde{\omega}}$ into the IIA action~\eqref{eq:siia} and expand with respect to large $\tilde{\omega}$\,, we find
\begin{align} \label{eq:expsiia}
	S^{}_\text{\scalebox{0.8}{IIA}} = \tilde{\omega} \, S^{}_\text{\scalebox{0.8}{NL}} + O\bigl( \tilde{\omega}^{-1} \bigr)\,,
\end{align}
where $S^{}_\text{NL}$ is the leading order action that we will focus on, with `NL' standing for `non-Lorentzian'. See \emph{e.g.}~\cite{VandenBleeken:2017rij, Hansen:2019pkl, Hansen:2020pqs, Hansen:2020wqw} for more general discussions on a covariant expansion of gravity with respect to large speed of light, where it is shown that extra field components should be introduced to understand the underlying symmetry algebra at higher orders. From this study, an action principle for Newtonian gravity was derived~\cite{Hansen:2019pkl}.\,\footnote{See also~\cite{Hartong:2021ekg, Hartong:2022dsx, Hartong:2024ydv} for an analogous expansion of string theory, in connection to our discussion on non-relativistic string theory in Appendix~\ref{app:pnrst}.}

Several definitions for various geometric quantities are in order before we present the explicit form of the truncated action $S^{}_\text{NL}$\,. In this truncated action, \emph{no} ten-dimensional metric description is available. Instead, a Newton-Cartan formalism in terms of the temporal vielbein $\tau^{}_\mu$ and spatial vielbein $e^{}_\mu{}^i$ is required, as the target space geometry develops a codimension-one foliation structure.\,\footnote{Strictly speaking, there is only a foliation structure when the twistless torsion constraint $\tau \wedge \dd \tau = 0$ is satisfied. Otherwise, we have a `distribution' rather than foliation.} We further define the inverse vielbein fields $\tau^\mu$ and $e^\mu{}_i$ via the orthogonality conditions,
\begin{subequations}
\begin{align}
	\tau^\mu_{} \, \tau^{}_\mu &= 1\,,
		&%
	\tau^\mu \, e^{}_\mu{}^i = e^\mu{}_i \, \tau^{}_\mu &= 0\,, \\[4pt]
	e^\mu{}_i \, e^{}_\mu{}^j &= \delta_i^j\,,
		&%
	\tau^\mu \, \tau_\nu + e^\mu{}_i \, e^{}_\nu{}^i &= \delta_\mu^\nu\,. 
\end{align}
\end{subequations}
We also introduce the spatial metrics 
$e^{}_{\mu\nu} \equiv e^{}_\mu{}^i \, e^{}_\nu{}^i$ and $e^{\mu\nu} \equiv e^\mu{}^{}_i \, e^\nu{}^{}_i$\,. 
Note that both the metrics are of rank 9 and thus degenerate, which means that there is \emph{no} inverse for either of them. 
In order to understand how to construct the covariant objects in the Newton-Cartan formalism, we present the symmetries of the truncated system. 
At the leading order, these vielbein fields and the RR 1-form $c^{(1)}$ acquire the Galilei boost transformations,
\be \label{eq:gbtec}
	\tau \rightarrow \tau + O\bigl(\tilde{\omega}^{-2}\bigr)\,,
		\quad%
	e^i \rightarrow e^i + v^i \, \tau + O\bigl(\tilde{\omega}^{-2}\bigr)\,,
		\quad%
	c^{(1)} \rightarrow c^{(1)} - e^{-\varphi} \, v^{}_i \, e^i + O\bigl(\tilde{\omega}^{-2}\bigr)\,,
\ee
where $v^i $ is the Galilei boost velocity. 
Moreover, $\tau \equiv \tau^{}_\mu \, \dd x^\mu$ and $e^i \equiv e^{}_\mu{}^i \, \dd x^\mu$. Note that $\tau^{}_\mu$ and $e^\mu{}^{}_i$ are boost invariant in the truncated theory. The SO(9) rotation symmetry on the spatial hypersurface is realized by contracting the frame indices $i$ in $e^i$ and the diffeomorphism invariance is realized by contracting the curved indices $\mu$. Moreover, we will see that $S_\text{\scalebox{0.8}{NL}}$ enjoys an extra emergent dilatation symmetry acting on the vielbein fields $\tau$ and $e^i$ as well as the dilaton $\varphi$ as
\be \label{eq:ds}
	\tau \rightarrow \Omega^{1/2} \, \tau\,,
		\qquad%
	e^i \rightarrow \Omega^{-1/2} \, e^i,
		\qquad%
	e^\varphi \rightarrow \Omega^{-3/2} \, e^\varphi,
\ee
where $\Omega$ is an arbitrary function. Further explanation of the origin of the dilatation symmetry will be given later in Section~\ref{sec:enls}. 

We are now ready to introduce the covariant quantities to describe $S_\text{\scalebox{0.8}{NL}}$. We first covariantize all derivatives with respect to the dilatation symmetry~\eqref{eq:ds}. For any operator $\CO$, we define the dilatation covariant derivative $\dd_\mu$ that acts on $\CO$ as
\be \label{eq:dmo}
	\dd_\mu \CO = \Bigl[ \p_\mu + \tfrac{2}{3} \, \Delta(\CO) \, \p_\mu \varphi \Bigr] \, \CO\,,
\ee  
where $\Delta(\CO)$ denotes the dilatation weight of the operator $\CO$. For examples, $\Delta(\tau) = \frac{1}{2}$ and $\Delta(e^i) = - \frac{1}{2}$\,. The covariant derivative $\dd_\mu$ in Eq.~\eqref{eq:dmo} is defined such that, under the transformation~\eqref{eq:ds}, both $\CO$ and $\dd_\mu \CO$ transform in the same way as
\be
    \CO \rightarrow \Omega^{\Delta(\CO)},
        \qquad%
    \dd_\mu \CO \rightarrow \Omega^{\Delta(\CO)} \, \dd_\mu \CO\,. 
\ee
In terms of $\dd_\mu$\,, any dilatation invariant object should not have any explicit $\varphi$ dependence other than an overall factor. It is understood that all $\p^{}_\mu$'s in the standard definitions are replaced with $\dd^{}_\mu$ in the following discussions, unless explicitly stated. Next, we introduce the compatibility conditions for the boost invariant quantities $\tau^{}_\mu$ and $e^\mu{}^{}_i$ in the truncated theory, 
\be
	\nabla^{}_{\!\mu} \tau^{}_\nu \equiv \dd^{}_\mu \tau^{}_\nu - \gamma^\rho{}^{}_{\mu\nu} \, \tau^{}_\rho = 0\,,
		\qquad%
	\nabla^{}_{\!\mu} e^{\nu\rho} \equiv \dd^{}_\mu e^{\nu\rho} + \gamma^\nu{}^{}_{\mu\sigma} \, e^{\sigma\rho} + \gamma^\rho{}^{}_{\mu\sigma} \, e^{\sigma\nu} = 0\,,
\ee
with $\gamma^\mu{}_{\rho\sigma}$ the Newton-Cartan connection, which we choose to be
\begin{align}
	\gamma^\mu{}^{}_{\rho\sigma} &= \tau^\mu \, \dd^{}_\rho \tau^{}_\sigma + \tfrac{1}{2} \, e^{\mu\nu} \, \bigl( \dd^{}_\rho e^{}_{\sigma\nu} + \dd^{}_\sigma e^{}_{\rho\nu} - \dd^{}_\nu e^{}_{\rho\sigma} \bigr)\,.
\end{align}
Note that $\gamma^\mu{}^{}_{\rho\sigma}$ contains an antisymmetric part that constitutes the torsion,
\be \label{eq:ts}
	t^{}_{\mu\nu} \equiv \tau^{}_\rho \, \gamma^{\rho}{}^{}_{[\mu\nu]} = \dd^{}_{[\mu} \tau^{}_{\nu]}\,. 
\ee
The associated Riemann tensor is
\be
	r^\rho{}_{\sigma\mu\nu} = \dd_\mu \gamma^\rho{}_{\nu\sigma} - \dd_\nu \gamma^\rho{}_{\mu\sigma} + \gamma^\rho{}_{\mu\lambda} \, \gamma^\lambda{}_{\nu\sigma} -\gamma^\rho{}_{\nu\lambda} \, \gamma^\lambda{}_{\mu\sigma}\,.
\ee
See \emph{e.g.}~\cite{Hartong:2022lsy} for extensive discussions on the relevant Newton-Cartan formalism. 

Finally, plugging Eq.~\eqref{eq:ftoa1} into the IIA action~\eqref{eq:siia}, we find that $S_\text{\scalebox{0.8}{NL}}$ in Eq.~\eqref{eq:expsiia} is
\begin{align} \label{eq:nlg}
    S^{}_\text{\scalebox{0.8}{NL}} &= \frac{1}{2 \, \tilde{\kappa}^2} \int \dd^{10} x \, e \, e^{-2\varphi} \, e^{\mu\nu} \, \bigl( r^{}_{\mu\nu} - 4 \, \tau^\rho \, \nabla^{}_{\!\mu} t^{}_{\nu\rho} - 6 \, \tau^\rho \, \tau^\sigma \, t^{}_{\mu\rho} \, t^{}_{\nu\sigma} - e^{\rho\sigma} \, e^\varphi \, t^{}_{\mu\rho} \, f^{}_{\nu\sigma} \Bigr)\,,
\end{align}
with $\tilde{\kappa}^2 \sim \tilde{\ell}_\text{s}^{\,8}$\,, $e \equiv \det (\tau_\mu\,,\,e_\mu{}^i)$\,, and $f_{\mu\nu} = \p_\mu c_\nu - \p_\nu c_\mu$\,. This action matches the one in~\cite{Lambert:2024ncn} up to a boundary term. 
Note that we have dropped a boundary term,
\be
    S^{}_\text{b} = - \frac{7}{3 \, \tilde{\kappa}^2} \int \dd^{10} x \, \p^{}_\mu \bigl( e \, e^{-2 \, \varphi} \, e^{\mu}{}^{}_i \, \p^{}_i \varphi \bigr)\,.
\ee
It is straightforward to check that $S^{}_\text{NL}$ is invariant under both the Galilei boost transformations~\eqref{eq:gbtec} (without the subleading terms) and dilatation transformations~\eqref{eq:ds}. Similar constructions of such a BPS decoupling limit of gravitational theories have also been studied in different contexts~\cite{Hartong:2022lsy, Bergshoeff:2024ipq}. 

We thus find that the leading-order contribution at large $\tilde{\omega}$ is described by the non-Lorentzian gravity~\eqref{eq:nlg}, expanded around the background geometry specified around~\eqref{eq:ftoa1}, with the non-trivial vacuum expectation values
$\langle \tau \rangle = \dd t / \tilde{h}^{1/2}$\,, 
$\langle e^i \rangle = \tilde{h}^{1/2} \, \dd x^i$\,, and
$\langle e^\varphi \rangle = \tilde{g}^{}_\text{s} \, \tilde{h}^{3/4}$. Nevertheless, performing the dilatation transformation~\eqref{eq:ds} with $\Omega = \tilde{h}$\,, we find $\langle \tau \rangle = \dd t$\,, $\langle e^i \rangle = \dd x^i$, and $\langle e^\varphi \rangle = \tilde{g}^{}_\text{s}$\,, which is the flat solution! We hence conclude that the IIA supergravity on the background \eqref{eq:rsnhg0} at the leading $O(\tilde{\omega})$ order behaves as a non-Lorentzian gravity around flat background. But one has to keep in mind that the non-Lorentzian supergravity only arises as a truncation of the full theory, which means that it can at most capture part of the modes in the BFSS matrix theory.  

\subsection{BFSS Matrix Theory at Moderately Large \texorpdfstring{$N$}{N}} \label{sec:mln}

We now turn our attention to the field theory side, which in the weakly coupled regime is described by a quantum mechanical system describing $N$ non-relativistic D-particles, \emph{i.e.}~the BFSS matrix theory. This is (0+1)-dimensional U($N$) super Yang-Mills (SYM) with 16 supercharges~\cite{Baake:1984ie, Flume:1984mn, Claudson:1984th},
\begin{equation} \label{eq:BFSSAction1}
    S_{\text{\scalebox{0.8}{BFSS}}} \!=\! \frac{1}{g_\text{s} \ell_\text{s}} \! \int \! \dd t\,\Tr\!\ls \tfrac{1}{2} \, D_\tau X^i \, D_\tau X^i + \frac{\ell_\text{s}^{-4}}{4} \bigl[X^i, X^j\bigr]^2 + \tfrac{1}{2} \, \psi^\intercal \Bigl( i \, D_\tau \psi + \ell_\text{s}^{-2} \, \gamma_i \, \bigl[ X^i,\psi \bigr] \Bigr) \rs\!. 
\end{equation}
Here, $X^i$ are nine $N \times N$ Hermitian matrices with $i = 1\,,\cdots,9$\,, $\psi$ is an $N \times N$ matrix-valued Majorana fermion in the $\mathbf{16}$ of Spin(9), $\gamma^i$ are the Dirac gamma matrices, and the covariant derivative acts as
\be
    D_\tau X^i = \p_\tau X^i - \! \bigl[A_\tau\,,X^i\bigr]\,,
        \qquad%
    D_\tau \psi = \p_\tau \psi - \! \bigl[A_\tau\,,\psi\bigr]\,,
\ee
with $A_\tau$ the only component of the U($N$) YM field. Focusing on the bosonic sector, the classical potential $[X^i,X^j]^2$ is minimized when $X^i$ is diagonalized, with\,\footnote{The quantum-mechanical ground state of the BFSS matrix theory is only known for specific limits~\cite{Lin:2014groundstate, Sethi:2000invariance, Porrati:1997boundstates, Konechny:1998hamiltonian, Moore:1998boundstates, Sethi:1997redux, Yi:1997threshold}.}
\begin{equation}
\bigl\langle X^i(t) \bigr\rangle = \begin{pmatrix}
x_{1}^i \, \mathbb{1}_{N_1} & 0 & 0 & 0 \\
 0 & x_2^i \, \mathbb{1}_{N_2} & 0 & 0 \\
 0 & 0 & \ddots & 0 \\
 0 & 0 & 0 & x_a^i \, \mathbb{1}_{N_a}
\end{pmatrix}\!,
\end{equation}
which describes a system of D-particle bound states located at $x_r^i$\,, $r = 1, \cdots, x_a^i$\,. At the position $x_r^i$\,, there is a bound state consisting of $N_r$ coinciding D-particles. Very na\"{i}vely, let us first directly apply the rescaling~\eqref{eq:gslsrs0} to the quantum mechanical action~\eqref{eq:BFSSAction1} and then expand with respect to large $\tilde{\omega}$, we find 
\begin{align} \label{eq:bfssexp}
\begin{split}
    S_{\text{\scalebox{0.8}{BFSS}}} = & \frac{\tilde{\omega}}{4 \, \tilde{g}_\text{s} \, \tilde{\ell}^{\,5}_\text{s}} \! \int \! \dd t\,\Tr \bigl[X^i, X^j\bigr]^2 + \frac{1}{2 \, \tilde{g}_\text{s} \, \tilde{\ell}^{\,3}_\text{s}} \int \dd t\, \Tr \Bigl( \psi^\intercal \, \gamma_i \, \bigl[ X^i,\psi \bigr] \Bigr) \\[4pt]
    	& + \frac{\tilde{\omega}^{-1}}{2 \, \tilde{g}_\text{s} \, \tilde{\ell}_\text{s}} \! \int \! \dd t\,\Tr \Bigl( D_\tau X^i \, D_\tau X^i 
	+ i \, \psi^\intercal \, D_\tau \psi \Bigr).
\end{split}
\end{align}
The kinematical contents now only appear at a subleading $O(\tilde{\omega}^{-1})$ order. The leading-order equation of motion from varying the action with respect to $X^i$ is
$[X^i, X^j] \, X^j = 0$\,,
which is solved by a diagonalized $X^i$. 
The bosonic sector of the BFSS matrix theory now reduces to  
\be \label{eq:sb}
	S^{}_\text{b} = \frac{N}{2 \, \tilde{\omega} \, \tilde{g}^{}_\text{s} \, \tilde{\ell}^{}_\text{s}} \int \dd t \, \p^{}_\tau X^i \, \p^{}_\tau X^i = \frac{N}{2 \, R^{}_\text{s}} \int \dd t \, \p^{}_\tau X^i \, \p^{}_\tau X^i, 
\ee
where the we have gauge fixed $A_\tau$ to zero. This action describes $N$ non-relativistic particles, with 
$R^{}_\text{s} = g^{}_\text{s} \, \ell^{}_\text{s} = \tilde{\omega} \, \tilde{g}^{}_\text{s} \, \tilde{\ell}^{}_\text{s}$
the radius of the M-theory circle. At large $\tilde{\omega}$\,, we enter the M-theory regime, and the D-particle action~\eqref{eq:sb} corresponds to a supergraviton with momentum $N / R_\text{s}$\,. At large $\tilde{\omega}$\,, together with a `moderately large' $N$ such that 
\be \label{eq:mln}
	\text{large } N,
		\qquad%
	\text{large } R^{}_\text{s}\,,
		\qquad%
	\frac{N}{R_\text{s}} \sim 0\,,
\ee
we zoom in on the low momentum sector. Note that we have seen a similar condition that requires $N$ be only moderately large on the bulk side below Eq.~\eqref{eq:rsnhg0}, such that $1 \ll N \ll \tilde{\omega}^2$\,. 
The conditions in Eq.~\eqref{eq:mln} are associated with the null-reduction of eleven-dimensional supergravity as the Kaluza-Klein momentum $N / R_\text{s}$ is close to zero. As shown in Eq.~\eqref{eq:sb}, the D-particle dynamics only enters at a subleading order $O\bigl( \tilde{\omega}^{-1} \bigr)$\,. In the next section, we will show that the null-reduction of eleven-dimensional supergravity precisely matches the non-Lorentzian supergravity on a flat background as discussed in Section~\ref{sec:nlse}. Although quantum mechanically the null reduction is typically \emph{not} self-contained~\cite{Hellerman:1997yu, Chapman:2020vtn}, we are saved by staying in a purely classical regime where the quantum corrections are suppressed, as the D-particles are decoupled at the leading order. 

Moreover, it follows immediately that the decoupled D-particles should not backreact in this non-Lorentzian supergravity. It is however worthwhile to explicitly show this by generalizing the above discussions to curved spacetime. 
We start with the coupling between a bound state of $N$ non-interacting D-particles and type IIA supergravity. This sourcing gives rise to the D-particle geometry as discussed in Section~\ref{sec:dph}. The $\tilde{\omega}$-expansion of the supergravity sector has been studied in Section~\ref{sec:nlse}, where we found that the leading $\tilde{\omega}$-order contribution is described by non-Lorentzian supergravity. Generalizing the non-relativistic D-particle action~\eqref{eq:sb} to include curved background fields, we write
\be \label{eq:dpsto}
	S^{}_\text{source} = \frac{N}{\tilde{\omega}} \lr \frac{1}{2} \int \dd \tau \, e^{-\varphi} \, \frac{e^{}_{\mu\nu} \, \dot{X}^\mu \, \dot{X}^\nu}{\tau^{}_\rho \, \dot{X}^\rho} + \int c^{(1)} \rr + O\bigl(\tilde{\omega}^{-3}\bigr). 
\ee
In particular, the static source arises from varying the action with respect to the RR one-form $c^{(1)}$ and then projecting using the temporal vielbein field. 
Under the condition $N \ll \tilde{\omega}^2$\,, this D-particle source is at a subleading order when compared with the supergravity expansion~\eqref{eq:expsiia}, where the leading-order non-Lorentzian supergravity appears to be at the order $O(\tilde{\omega})$\,. Therefore, there is \emph{no} backreaction of D-particles in this non-Lorentzian supergravity, which explains why the background geometry is flat, as we have seen at the end of Section~\ref{sec:nlse}. Furthermore, this implies that the D-particles do \emph{not} interact at a distance. This is accordance with that, in the strong coupling regime, off-diagonal
modes of the matrices $X^i$ decouple. These modes correspond to open strings stretched between
the D-particles and mediating long-range interactions. Therefore, interactions involving the D-particles can only
arise from collisions of localized D0-brane bound states, \emph{i.e.}~when their worldlines intersect.

Next, we consider the backreaction of extended objects such as the F-string and D$p$-branes with $p \neq 0$\,, focusing on the brane tension sector without gauge potentials. Using $\sigma^\alpha$, $\alpha = 0\,,\,\cdots,\,p$ to denote the coordinates on the worldvolume, we write the F-string and D$p$-brane actions in general background fields as 
\begin{subequations} \label{eq:sba}
\begin{align}
    S^{}_\text{\scalebox{0.8}{F1}} &= - \frac{1}{\ell_\text{s}^2} \int \dd^2 \sigma \, \sqrt{-\det G^{}_{\alpha\beta}} - \frac{1}{\ell_\text{s}^2} \int B^{(2)}, \label{eq:fswsa} \\[4pt]
	S^{}_\text{\scalebox{0.8}{D$p$}} &= - \frac{1}{\ell_\text{s}^{p+1}} \int \dd^{p+1} \sigma \, e^{-\Phi} \sqrt{-\det\Bigl( G^{}_{\alpha\beta} + \CF^{}_{\alpha\beta} \Bigr)} + \frac{1}{\ell_\text{s}^{p+1}} \int \sum_{q} C^{(q)} \wedge e^{\CF^{(2)}}\bigg|_{p+1}, \label{eq:dpbwa}
\end{align}
\end{subequations}
where $\CF^{(2)} = B^{(2)} + F^{(2)}$, with $G_{\mu\nu}$ the metric field, $B^{(2)}$ the Kalb-Ramond field, $C^{(q)}$ the RR fields, and $F^{(2)}$ the $U(1)$ gauge field strength on the D$p$-brane. It is understood that $G^{}_{\alpha\beta} = \p^{}_\alpha X^\mu \, \p^{}_\beta X^\nu \, G_{\mu\nu}$ and $B^{}_{\alpha\beta} = \p^{}_\alpha X^\mu \, \p^{}_\beta X^\nu \, B_{\mu\nu}$ are pullbacks. Focusing on the metric and dilaton part and setting the gauge potentials to zero, we use Eq.~\eqref{eq:ftoa1} and $\ell_\text{s} = \tilde{\ell}_\text{s} / \sqrt{\tilde{\omega}}$ to expand the worldvolume actions in Eq.~\eqref{eq:sba} with respect to large $\tilde{\omega}$ as
\begin{subequations} \label{eq:efodp}
\begin{align}
    S^{}_{\text{\scalebox{0.8}{F1}}} &= - \frac{\tilde{\omega} \, N}{\tilde{\ell}^{\,2}_\text{s}} \int \dd^2 \sigma \, \sqrt{- \det \CM_3} + O\bigl(\tilde{\omega}^{-1}\bigr)\,, \label{eq:fos} \\[4pt]
	S^{}_{\text{\scalebox{0.8}{D$p$}}} &= - \frac{\tilde{\omega} \, N}{\tilde{\ell}^{\,p+1}_\text{s}} \int \dd^{p+1} \sigma \, e^{-\varphi} \sqrt{- \det 
		\CM_{p+2}} + O\bigl(\tilde{\omega}^{-1}\bigr)\,, 
            \qquad%
        p \neq 0\,,
\end{align}
\end{subequations}
where $\sigma^\alpha$, $\alpha = 0\,,\,\cdots,\,p$ are the worldvolume coordinates and $\CM_n$ is an $n \times n$ matrix, with 
\be \label{eq:sdps}
    \CM_{p+2} = 
    \begin{pmatrix}
		0 &\,\,\,\, \p^{}_\beta X^\nu \, \tau^{}_\nu \\[4pt]
		\p^{}_\alpha X^\mu \, \tau^{}_\mu &\,\,\,\, \p^{}_\alpha X^\rho \, \p^{}_\beta X^\sigma e^{}_{\rho\sigma}
	\end{pmatrix},
        \qquad%
    \alpha = 0\,,\,\cdots,\,p\,.
\ee
Similarly, the tension part of the NS5-brane action is
\begin{subequations} \label{eq:nsf}
\begin{align} 
    S^{}_\text{\scalebox{0.8}{NS5}} &= - \frac{\tilde{\omega} \, N}{\tilde{\ell}^{\,6} _\text{s}} \int \dd^6 \sigma \, e^{-2 \, \varphi} \, \sqrt{- \det \CM_{7}} + O\bigl(\tilde{\omega}^{-1}\bigr)\,.
\end{align}
\end{subequations}
All these actions contain a leading $\tilde{\omega}$ piece that potentially sources the non-Lorentzian supergravity in Section~\ref{sec:nlse}. This implies that the truncated non-Lorentzian supergravity may still contain interesting dynamics and non-trivial curved solutions, from backreacting the strings or branes other than the D-particles. 
Understanding how static strings and branes are coupled to the non-Lorentzian supergravity requires a better knowledge of the Kalb-Ramond and RR sector in the large $\tilde{\omega}$ expansion, which we will return to in Section~\ref{sec:bth} after discussing the DLCQ of M-theory in the next section. We will see that, at $O(\tilde{\omega})$\,, only the D4-brane and F1-string acquire a charge term. A necessary condition comes from the BPS nature of these configurations: recall that the large $\tilde{\omega}$ expansion is performed with respect to a background D-particle, which can form a $\frac{1}{4}$-BPS state with either the D4-brane or F1-string. In later sections we will discuss the associated curved non-Lorentzian geometries as solutions to the truncated non-Lorentzian supergravity.

The above discussion 
also extends beyond the regime $N^{1/7} \ll r / \ell^{}_{11} \ll N^{1/3}$. Within the 11-dimensional supergravity regime $N^{1/9} \ll r / \ell^{}_{11} \lesssim N^{1/7}$, under the $\tilde{\omega}$ reparametrization the pp-wave metric~\eqref{eq:ppw} becomes
\be \label{eq:ppw2}
	\dd s^2_{11} = 2 \, \dd t \, \dd u + \frac{\tilde{h}}{\tilde{\omega}^{2}} \, \dd u^2 + \dd r^2 + r^2 \, \dd \Omega^2_8\,,
\ee 
and the leading-order contribution is the null-reduced geometry
$\dd s^2_{11} = 2 \, \dd t \, \dd u + \dd r^2 + r^2 \, \dd \Omega^2_8$\,, 
where $u$ is compactified.

In the next section we will show that applying the decoupling limit to IIA supergravity without D-particle backreaction leads to a non-Lorentzian supergravity describing the instantaneous interactions, deprived of asymptotic gravitational states~\cite{Blair:2023noj, Gomis:2023eav, Blair:2024aqz}. 
Later in Section~\ref{sec:bth} we will show that further taking the $\tilde{\omega}$ expansion introduced in the current section and the truncation at the leading order decouples the dynamics of the D-particles and thus isolates the classical non-Lorentzian supergravity. This is the null reduction of the eleven-dimensional supergravity obtained from the moderately large $N$ expansion in Eq.~\eqref{eq:mln}.

\section{Non-Lorentzian Limit of String Theory} \label{eq:nllst}

Let us now return to the decoupling limit introduced in Eq.~\eqref{eq:ftl}, which zooms in on the D-particles and leads to the BFSS matrix theory. This decoupling limit is closely related to the discrete light cone quantization (DLCQ) of M-theory, \emph{i.e.}~M-theory compactified over a null circle~\cite{Susskind:1997cw, Sen:1997we, Seiberg:1997ad}. We consider the application of such a decoupling limit to the full type IIA superstring theory, which leads to a corner referred to as Matrix 0-brane Theory (M0T) in~\cite{Blair:2023noj, Gomis:2023eav, Blair:2024aqz}. This is a fully quantum mechanical system whose low-energy classical description is captured by a non-Lorentzian supergravity, which arises from the same decoupling limit but now applied to IIA supergravity. This non-Lorentzian supergravity corresponds to the null reduction of eleven-dimensional supergravity, which we will later use to complete the analyses in Section~\ref{sec:nlrdph}. After reviewing the basics of DLCQ M-theory and M0T, we will study the associated non-Lorentzian supergravity and analyze its dynamics. We will also argue that this dynamics should be captured by a chiral string worldsheet theory akin to ambitwistor string theory~\cite{Mason:2013sva}. 

\subsection{Non-Lorentzian Quantum Gravity?}

It was argued in~\cite{Susskind:1997cw, Sen:1997we, Seiberg:1997ad} that the finite-$N$ BFSS matrix theory should describe the dynamics of Kaluza-Klein excitations of eleven-dimensional M-theory over a lightlike compactification, \emph{i.e.}~the DLCQ of M-theory. Therefore, the correlation functions in the perturbative BFSS matrix theory should correspond to certain scattering processes in the DLCQ of eleven-dimensional supergravity. But the physics associated with such a lightlike compactification appears to be exotic and it is not uncommon to doubt whether DLCQ is even well defined. Following the same logic of how M-theory is related to type IIA superstring theory, we know that the perturbative BFSS matrix theory corresponds to M-theory on a small lightlike circle. But what is eleven-dimensional supergravity on a vanishing lightlike circle? This might appear to be even more singular than the DLCQ itself. 

More recently, based on new developments of non-Lorentzian geometric techniques, it was demonstrated in~\cite{Bergshoeff:2018yvt} that the DLCQ of string theory receives a first principles definition from non-relativistic string theory~\cite{Klebanov:2000pp, Gomis:2000bd, Danielsson:2000gi}. This string theory arises from a well-defined BPS decoupling limit zooming in on a background fundamental string: namely, the string charge is fine tuned to cancel an infinite string tension. Such a decoupling limit breaks the target space Lorentz symmetry spontaneously. As a result, non-relativistic string theory has a Galilei-like target space with a codimension-two foliation structure~\cite{Andringa:2012uz}, where wound strings interact with each other via an instantaneous Newton-like gravitational force~\cite{Gomis:2000bd, Danielsson:2000mu}. Since the fundamental string worldsheet in this string theory remains relativistic, standard worldsheet conformal field theoretical techniques continue to apply, and the target space physics can be derived from the string worldsheet perspective as usual. For example, the beta functions of the worldsheet sigma models describing the non-relativistic string are studied in~\cite{Gomis:2019zyu, Yan:2019xsf, Gallegos:2019icg, Yan:2021lbe}, which shows that perturbative non-relativistic string theory describes a quantum gravity with an associated effective geometry that is non-Lorentzian. It has been demonstrated in~\cite{Bergshoeff:2018yvt, Gomis:2020izd} that T-dualizing non-relativistic string theory along a spatial circle longitudinal to the background fundamental string defines the DLCQ of relativistic string theory. Similarly, the lightlike circle in DLCQ M-theory is mapped to a spatial torus in a U-dual frame, where the spacetime acquires a non-Lorentzian structure adapted to a BPS decoupling limit zooming on a background membrane~\cite{Ebert:2023hba, Blair:2024aqz}. 

On the other hand, the second quantization of non-relativistic string theory is matrix string theory, which is an orbifold CFT described by a two-dimensional $\CN = 8$ SYM and is dual to the BFSS matrix theory compactified on a spatial circle~\cite{Motl:1997th, Dijkgraaf:1997vv}. Since the F-strings in non-relativistic string theory are dual to the D-particles in the BFSS matrix theory, one expects that the latter provide a worldline (instead of worldsheet) description of the target space physics. At finite $N$, this is a ten-dimensional quantum gravity associated with an effective non-Lorentzian geometry, in which the time direction is absolute.  

When $N$ becomes large, the perturbative description of the BFSS matrix theory and its corresponding non-Lorentzian quantum gravity are no longer valid. This is because there are now a large number of D-particles whose backreaction creates a Lorentzian bulk geometry. This non-perturbative process underlies general holographic duals in string theory including the AdS/CFT correspondence. A modern treatment of the relations between BPS decoupling limits, non-Lorentzian geometry, holography, and the $T\bar{T}$ deformation is recently given in~\cite{Blair:2024aqz}.  

\subsection{From BPS Decoupling Limit to DLCQ M-Theory}

Before studying the non-Lorentzian supergravity that arises in the D-particle decoupling limit, we first review some essential ingredients of DLCQ M-theory~\cite{Susskind:1997cw, Sen:1997we, Seiberg:1997ad}. In order to motivate the M-theory picture, it is helpful to start with the non-relativistic limit of the D-particle worldline theory.  

\vspace{3mm}

\noindent $\bullet$~\emph{Non-relativistic D-particles.} We start with examining the infinite speed of light limit of the D-particle in type IIA superstring theory. In flat spacetime, focusing on the bosonic sector, the action describing a D-particle coupled to the RR one-form $C^{(1)}$ is
\be
    S_\text{\scalebox{0.8}{D0}} = - m \, c \int \dd \tau \, \sqrt{c^2 \, \dot{X}_0^2 - \dot{X}^i \, \dot{X}^i} + \frac{c^2}{\ell_\text{s}} \int \dd \tau \, \dot{X}^\mu \, C^{}_\mu\,,
        \qquad%
    m = \bigl( g_\text{s} \, \ell_\text{s} \bigr)^{-1},
\ee
with $\tau$ the proper time and the dependence on the speed of light $c$ explicitly displayed. We take the static gauge with $X^0 = \tau$. In the strict infinite $c$ limit, the rest mass diverges, which we cancel by turning on a critical electric potential $C_0 = g_s^{-1}$. 
assuming that $\dot{X}^0 > 0$\,, which gives rise to the finite action,
\be \label{eq:nrpa}
    S_\text{non-rel.} = \frac{m}{2} \int \dd \tau \, \dot{X}^i \, \dot{X}^i\,.
\ee
This limiting procedure can be thought of as a $c \rightarrow \infty$ limit of type IIA superstring theory, with the metric $G_{\mu\nu}$, dilaton $\Phi$, and Ramond-Ramond one-form $C^{(1)}$ background fields reparametrized as
\begin{align} \label{eq:repo}
    G_{\mu\nu} \, \dd x^\mu \, \dd x^\nu = - c \, \dd t^2 + \frac{1}{c} \, \dd x_i^2\,,
        \qquad%
    e^\Phi = c^{-\frac{3}{2}} \, g^{}_\text{s}\,,
        \qquad%
    C^{(1)} = \frac{c^2}{g^{}_\text{s}} \, \dd t\,.
\end{align}
Together with a reparametrization of the fermionic field, $\Theta = c^\frac{1}{4} \, \Theta_- + c^{-\frac{3}{4}} \, \Theta_+$\,, and generalized to the case of a stack of $N$ coinciding D0-branes, the BFSS matrix theory~\eqref{eq:BFSSAction1} is recovered in the $c \rightarrow \infty$\,. See further discussions in~\cite{Gomis:2004pw, Blair:2024aqz}. 

Focusing on the reparametrizations of the bosonic background fields in Eq.~\eqref{eq:repo}, its uplift to M-theory is captured by the standard Kaluza-Klein reduction ansatz,
\begin{align} \label{eq:lcdlcq}
\begin{split}
    \dd s_{11}^2 &= l_\text{P}^{2} \, \Bigl[ e^{4\Phi/3} \bigl( R^{-1} \, \dd x^{10} + l_\text{s}^{-1} \, C^{(1)} \bigr)^{\!2} + e^{-2\Phi/3} \, l_\text{s}^{-2} \, G_{\mu\nu} \, \dd x^\mu \, \dd x^\nu \Bigr] \\[4pt]
    &= 2 \, \dd x^+ \, \dd x^- + \dd x^i \, \dd x^i + c^{-2} \, \dd x^- \, \dd x^-,
        \qquad%
    x^- \equiv x^{10},
        \qquad%
    x^+ \equiv t.
\end{split}
\end{align}
where $\ell^{}_\text{P} = g_\text{s}^{1/3} \, \ell^{}_\text{s}$ denotes the Planck length in M-theory and the dimensional reduction is taken in the eleventh dimension $x^- = x^{10}$, which is compactified over a circle of radius $R = g^{}_\text{s} \, \ell^{}_\text{s}$\,, with $x^{-} \sim x^{-} + 2 \pi R$\,. Taking the $c \rightarrow \infty$ limit in M-theory, we are led to
\be \label{eq:ds11}
    \dd s_{11}^2 \rightarrow 2 \, \dd x^+ \, \dd x^- + \dd x^i \, \dd x^i\,. 
\ee
Note that the non-compact direction $x^+$ corresponds to the time direction $t$ in from the string theory perspective. Moreover, the M-theory circle $x^-$ becomes lightlike in the infinite $c$ limit. This is M-theory in the DLCQ, \emph{i.e.}~M-theory compactified over the lightlike circle in $x^-$. 

\vspace{3mm}

\noindent $\bullet$~\emph{Infinite boost.} This infinite $c$ limit also receives a heuristic interpretation as an `infinite boost' limit in a compact, spacelike $y^{10}$ direction with radius $R_\text{s}$\,. However, since we are performing the infinite boosting in a compact direction, which already breaks Lorentz symmetry, this procedure should be thought of as a decoupling limit zooming in on a Kaluza-Klein state that is BPS, rather than a mere change of the reference frame. 

Consider the flat metric $\dd s^2 = - \dd y_0^2 + \dd y_{10}^2 + \dd x^i \, \dd x^i$ and a Lorentz boost in $y^{10}$ with the rapidity $\theta$. In terms of the light-cone coordinates $y^\pm = (y^{10} \pm y^0) / \sqrt{2}$, the boost transformation is $\tilde{y}^{\,\pm} = e^{\mp\theta} \, y^{\,\pm}$\,, where $\tilde{y}^{\,\pm}$ denote the boosted light-cone coordinates. In order to match on to the prescription~\eqref{eq:lcdlcq}, we further define
$x^+ = {\tilde{y}}^{\,+} - e^{-2\theta} \, {\tilde{y}}^{\,-}$ and $x^- = {\tilde{y}}^{\,-}$.
Now, the periodic boundary condition becomes
\be
	(x^+, x^-) \sim (x^+, x^- \! + 2\pi R)\,,
		\qquad%
	R = \omega \, R^{}_\text{s}\,,
        \qquad%
    \omega \equiv e^\theta / \sqrt{2}\,.
\ee
Upon identifying $c$ with 
$\omega$\,, the eleven-dimensional line element matches Eq.~\eqref{eq:lcdlcq}. 
Note that, instead of the infinite speed of light limit controlled by the dimensionful parameter $c$ (measured in spatial and temporal units), it is more appropriate to treat the BPS decoupling limit as controlled by the dimensionless quantity $\omega$ as in Eq.~\eqref{eq:dlo} associated with the infinite boost in eleven dimensions.  We will therefore stick to the infinite $\omega$ limit instead of the infinite $c$ limit in the rest of the paper.

\vspace{3mm}

\noindent $\bullet$~\emph{BFSS conjecture.} We now review what the infinite boost limit implies for the dynamics. We start with the M-theory geometry~\eqref{eq:lcdlcq}, but now in terms of the parameter $\omega$, such that
\be
    \dd s^2_{11} = 2 \, \dd x^+ \, \dd x^- + \dd x^i \, \dd x^i + \omega^{-2} \, \dd x^- \, \dd x^-,
        \qquad%
    x^- \sim x^- \! + 2 \pi R\,.
\ee
The M-theory uplift of a D-particle bound state with mass $N / (g_\text{s} \, \ell_\text{s})$ in type IIA superstring theory is the Kaluza-Klein state associated with a supergraviton with momentum $p^{}_-$ in the $x^-$ direction, which is a spacelike isometry. Since $x^-$ is compact, the momentum $p^{}_- = N / R$ is quantized, with $N \in \mathbb{Z}$ and $R = g_\text{s} \, \ell_\text{s}$\,. 

In the infinite $\omega$ limit, the $x^-$ circle becomes lightlike and we are led to DLCQ M-theory, whose dynamics is captured by bound states of non-relativistic D-particles, described by the BFSS matrix theory~\eqref{eq:BFSSAction1}. In the large $R$ limit, the lightlike circle decompactifies, which gives rise to eleven-dimensional M-theory. In order for the bound state mass $N / R$ to stay finite, a large $N$ limit is also simultaneously required. It is therefore conjectured that the large $N$ limit of the BFSS matrix theory should describe the full M-theory in asymptotically flat spacetime~\cite{Banks:1996vh}. In this limit, the BFSS effective coupling~\eqref{eq:geff} becomes infinitely large, which is highly non-perturbative. 

At large $N$, the supergraviton in eleven dimensions has a large null momentum, and its backreaction on the background geometry cannot be ignored. However, backreacting the supergraviton will deform the null isometry into a spacelike one, and the background geometry is now described by the Aichelburg-Sexl metric~\cite{Becker:1997xw}. We will return to this backreaction later in Section~\ref{sec:bth}. 

\subsection{Effective Non-Lorentzian Supergravity from a Limit} \label{sec:enls}

We now study the target space low-energy effective gravity related to the perturbative BFSS matrix theory at finite $N$, where backreaction of the D-particles is negligible. In this regime, a good approximation is given by the strict BPS decoupling limit~\eqref{eq:repo} of type IIA superstring theory. We transcribe such a limit in terms of the dimensionless parameter $\omega$ as~\cite{Blair:2023noj} 
\be \label{eq:m0tlp}
    \dd s^2_{10} = - \omega \, \dd t^2 + \frac{1}{\omega} \, \dd x^2_i\,,
        \qquad%
    e^\Phi = \omega^{-\frac{3}{2}} \, g^{}_\text{s}\,,
        \qquad%
    C^{(1)} = \frac{\omega^2}{g^{}_\text{s}} \, \dd t\,.
\ee
This $\omega \rightarrow \infty$ limit associated with the above reparametrization of the background fields is equivalent to the original decoupling limit prescribed by Eq.~\eqref{eq:dlo}, now with the rescaling of the string length absorbed into the reparametrizations of the background fields. We apply the $\omega \rightarrow \infty$ limit prescribed by Eq.~\eqref{eq:m0tlp} to the full type IIA superstring theory instead of only the D-particles, which leads to a corner of string theory that corresponds to dimensionally reducing DLCQ M-theory along the lightlike circle. We refer to this corner of the full type IIA superstring theory as \emph{Matrix 0-brane Theory} (M0T)~\cite{Blair:2023noj, Gomis:2023eav}. 
The target space geometry in M0T is non-Lorentzian, due to the BPS decoupling limit zooming in on a D-particle state, which spontaneously breaks the ten-dimensional Lorentz symmetry. This non-Lorentzian nature is made manifest by considering the probe D-particle described by the action~\eqref{eq:nrpa}, which enjoys a linear shift symmetry, $X^i \rightarrow X^i + v^i \, \tau$\,, which is the Galilei boost symmetry from the target space perspective.\,\footnote{It can be shown that, under a T-duality transformation, higher-dimensional matrix gauge theories also enjoy an analogous linear shift symmetry. This is akin to the Galileon symmetry that admits a braneworld origin~\cite{Nicolis:2008in}, which is a special case of polynomial shift symmetries~\cite{Griffin:2013dfa}.} 
In the low-energy regime, it is expected that M0T should acquire a classical non-Lorentzian supergravity description. It is in this sense that one could consider M0T as a non-Lorentzian quantum gravity.

So far we have been focusing on flat background. In order to study the low-energy non-Lorentzian supergravity, we first generalize to arbitrary curved background fields by introducing the vielbein fields that covariantize the one-forms $\dd t$ and $\dd x^i$, \emph{i.e.}~$\dd t \rightarrow \tau^{}_\mu \, \dd x^\mu$ and
$\dd x^i \rightarrow e^{}_\mu{}^i \, \dd x^\mu$.
Furthermore, define the dilaton, Kalb-Ramond, and RR fields in M0T to be $\varphi$, $b^{(2)}$, and $c^{(q)}$ (with $q$ odd), respectively. As in~\cite{Blair:2023noj, Gomis:2023eav}, we generalize the limiting prescription~\eqref{eq:m0tlp} to arbitrary background fields to be
\begin{subequations} \label{eq:m0tcb}
\begin{align}
    G^{}_{\mu\nu} &= - \omega \, \tau^{}_\mu \, \tau^{}_\nu + \omega^{-1} \, e^{}_\mu{}^i \, e^{}_\nu{}^i,
        &%
    C^{(1)} &= \omega^2 \, e^{-\varphi} \, \tau^{}_\mu \, \dd x^\mu + c^{(1)}, \\[4pt]
    B^{(2)} &= b^{(2)},
        \qquad%
    \Phi = \varphi - \tfrac{3}{2} \, \ln \omega\,,
        &%
    C^{(q)} &= c^{(q)}, \quad \text{odd }q \neq 1\,.
\end{align}
\end{subequations}
It is understood that the Hodge
dual relations between the RR potentials are taken into account. 
In the $\omega \rightarrow \infty$ limit, the ten-dimensional metric becomes singular. However, the vielbein formalism in terms of the temporal component $\tau^{}_\mu$ and spatial components $E^{}_\mu{}^i$ is still valid, which constitutes the non-Lorentzian geometry. 

In the $\omega\rightarrow\infty$ limit of the D-particle action in arbitrary background fields, the flat spacetime action~\eqref{eq:nrpa} is generalized to be
\be \label{eq:d0a}
    S^{\text{\scalebox{0.8}{M0T}}}_{\text{\scalebox{0.8}{D0}}} = \frac{1}{2 \, \ell_\text{s}} \int \dd \tau \, e^{-\varphi} \, \frac{\dot{X}^\mu \, \dot{X}^\nu \, e^{}_{\mu\nu}}{\dot{X}^\rho \, \tau^{}_\rho} + \int c^{(1)},
\ee
which matches the $\tilde{\omega}^{-1}$ order term in the action~\eqref{eq:dpsto}.
We now apply the same $\omega\rightarrow\infty$ limit to the bosonic sector of IIA supergravity. Our starting point is the supergravity action~\eqref{eq:siia} in string frame, but now extended to include the gauge potentials $B^{(2)}$ and $C^{(q)}$ with $q \neq 1$\,. The IIA action is
\begin{align} \label{eq:siiag}
	S^{}_{\text{\scalebox{0.8}{IIA}}} &= \frac{1}{2 \, \kappa^2} \int \dd^{10} x \, \sqrt{-G} \, \biggl[ e^{-2\Phi} \Bigl( R + 4 \, \p_\mu \Phi \, \p^\mu \Phi - \tfrac{1}{2} \bigl| H^{(3)} \bigr|^2 \Bigr) -\tfrac{1}{2} \, \Bigl( \bigl| F^{(2)} \bigr|^2 + \bigl| \tilde{F}^{(4)} \bigr|^2 \Bigr) \biggr] \notag \\[4pt]
	&\quad - \frac{1}{4 \, \kappa^2} \int B^{(2)} \wedge F^{(4)} \wedge F^{(4)}\,,
\end{align}
where $H^{(3)} \equiv \dd B^{(2)}$, $F^{(p+1)} \equiv \dd C^{(p)}$, and $\tilde{F}^{(4)} \equiv \dd C^{(3)} + C^{(1)} \wedge H^{(3)}$. Note that the coupling $\kappa^2 \sim \ell^{\,8}_\text{s}$ is independent of the string coupling constant $g^{}_\text{s}$\,. Moreover, for a $p$-form $\CT^{(p)}$, 
\be
	\bigl| \CT^{(p)} \bigr|^2 = \tfrac{1}{p!} \, G^{\mu^{}_1 \nu^{}_1} \, \cdots \, G^{\mu^{}_p \nu^{}_p} \, \CT^{}_{\mu_1 \cdots \mu_p} \, \CT^{}_{\nu_1 \cdots \nu_p}\,.
\ee
Plug the ansatz~\eqref{eq:m0tcb} into the IIA action~\eqref{eq:siiag} and up to a boundary term, we find in the $\omega \rightarrow \infty$ limit the following non-Lorentzian supergravity action previously studied in~\cite{Lambert:2024ncn}, which was shown to be equivalent to the null reduction of eleven-dimensional supergravity:
\begin{align} \label{eq:expsiia2}
	S^{}_\text{\scalebox{0.8}{IIA}} \rightarrow S^{}_\text{\scalebox{0.8}{NL}} &= \frac{1}{2 \, \kappa^2} \int \dd^{10} x \, e \, e^{-2\varphi} \, e^{\mu\nu} \, \bigl( r^{}_{\mu\nu} - 4 \, \tau^\rho \, \nabla^{}_{\!\mu} t^{}_{\nu\rho} - 6 \, \tau^\rho \, \tau^\sigma \, t^{}_{\mu\rho} \, t^{}_{\nu\sigma} - e^{\rho\sigma} \, e^\varphi \, t^{}_{\mu\rho} \, f^{}_{\nu\sigma} \Bigr) \notag \\[4pt]
	&\quad - \frac{1}{4 \, \kappa^2} \int \dd^{10} x \, e \, \Bigl( - \tfrac{1}{2} \, e^{-2 \, \varphi} \, h^{}_{0ij} \, h^{}_{0ij} + \tfrac{1}{3} \, e^{- \varphi} \, h^{}_{ijk} \, \tilde{f}^{}_{0ijk} + \tfrac{1}{4!} \, \tilde{f}^{}_{ijk\ell} \, \tilde{f}^{}_{ijk\ell} \Bigr) \notag \\[4pt]
	&\quad - \frac{1}{4 \, \kappa^2} \int b^{(2)} \wedge \tilde{f}^{(4)} \wedge \tilde{f}^{(4)},
\end{align}
with $e \equiv \det (\tau_\mu\,\,\,e_\mu{}^i)$\,, 
$h^{(3)} \equiv \dd b^{(2)}$, $f^{(2)} \equiv \dd c^{(1)}$, and $\tilde{f}^{(4)} \equiv \dd c^{(3)} +   c^{(1)} \wedge h^{(3)}$. Tensors with flat indices are understood to be projected by vielbein fields, \emph{e.g.},
\be
    h^{}_{0ij} \equiv \tau^\mu \, e^{\nu}{}^{}_i \, e^{\rho}{}^{}_j \, h^{}_{\mu\nu\rho}\,.
\ee
Note that $S^{}_\text{NL}$ is exactly the non-Lorentzian supergravity action in Eq.~\eqref{eq:expsiia}, but now it also includes non-zero higher-form gauge potentials. This action is invariant under the Galilei boost transformations,
\be
    \delta^{}_\text{\scalebox{0.8}{G}} \tau^{}_\text{} = 0\,,
		\qquad%
	\delta^{}_\text{\scalebox{0.8}{G}} e^i_\text{} = v^i \, \tau,
		\qquad%
	\delta^{}_\text{\scalebox{0.8}{G}} c^{(1)} = e^{-\varphi}\,v^i \, e^i,
\ee
as well as the emergent dilatation symmetry~\eqref{eq:ds}. This emergent local symmetry originates from that $\omega$ is set to infinitely large in Eq.~\eqref{eq:m0tcb}, and it removes one
of degrees of freedom of relativistic supergravity. Note that a total of 111 equations of motion arise from varying the action~\eqref{eq:expsiia2} with respect to $\tau_\mu$ (10 components), $e^{}_\mu{}^i$
(90 components), $\varphi$ (1 component), and $c^{}_\mu$ (10 components).
46 of these equations are Noether identities: 1 associated with dilatation symmetry, 9 Galilei boosts, and 36 spatial rotations.
This leaves us with 65 independent equations. In contrast, in the relativistic case, there are 45 Noether identities arising from
Lorentz symmetries. As a result, there are 66 independent equations of motion from
varying in $G^{}_{\mu\nu}$, $C_\mu$, and $\Phi$. This missing equation arises from varying the original IIA theory~\eqref{eq:siiag} with respect to $\omega$, which we now promote to be a dynamical field. This promotion does not affect the $\omega \rightarrow \infty$ limit. The same set of non-Lorentzian gravitational field equations can also be obtained from taking the $\omega \rightarrow \infty$ limit of the IIA supergravity equations of motion. This is in contrast to the expansion of supergravity as we have discussed in Section~\ref{sec:nlse}. There, the non-Lorentzian supergravity sector is a truncation of the full theory, and the associated equations of motion at this leading order do not contain such a missing equation associated with the dilatation transformation. This can also be understood by noting that the IIA action~\eqref{eq:expsiia} is not invariant under the dilatation transformation, only $S^{}_\text{\scalebox{0.8}{NL}}$ is. See Appendix~\ref{app:eom} for the the detailed expressions from varying the non-Lorentzian supergravity action with respect to the background fields. 

\subsection{Non-Lorentzian Gravity from String Worldsheet?} \label{sec:nlgsw}

It is interesting to understand whether there is a string theoretic origin of the non-Lorentzian gravitational dynamics captured by the action~\eqref{eq:expsiia2}. In particular, in the dual frame of M0T compactified over a spatial circle described by non-relativistic string theory, the standard beta-function calculation becomes available. This is because the fundamental degrees of freedom in non-relativistic string theory are the F-strings, which are dual to be the D-particles in M0T. While the dynamics of the D-particles in M0T is described by the BFSS matrix theory, the dynamics of the F-strings in non-relativistic string theory is described by matrix string theory. One may therefore expect that there is a dual role of the beta function calculation in M0T. An idea pointing towards such a possibility comes from ambitwistor string theory~\cite{Mason:2013sva}, where it is still the string worldsheet theory that encodes the (non-Lorentzian) gravitational equations of motion. 

Instead of focusing on M0T, we first consider its T-duality transformation along a timelike isometry~\cite{Hull:1998vg}. Instead of being an equivalent T-dual description, this transformation should be thought of as an analytic continuation. Such a timelike T-dual transformation maps the limiting prescription~\eqref{eq:m0tcb} to be~\cite{Blair:2023noj}
\begin{subequations} \label{eq:mm1tcb}
\begin{align}
    \dd s^2_{10} & = \frac{e^{}_{\mu\nu}}{\omega} \, \dd x^\mu \, \dd x^\nu,
        &%
    C^{(0)} &= \frac{\omega^2}{e^\varphi} + c^{(0)}, \\[4pt]
    B^{(2)} &= b^{(2)},
        \qquad%
    e^\Phi = \frac{e^\varphi}{i \, \omega^2}\,, 
        &%
    C^{(2)} &= c^{(2)}, 
    	\qquad%
    C^{(4)} = c^{(4)}.
\end{align}
\end{subequations}
Here, $e_{\mu\nu} = e_\mu{}^a \, e_\nu{}^b \, \eta_{ab}$ with $a = 0\,,\,\cdots,\,9$ is a ten-dimensional metric and the target space geometry is Lorentzian. The presence of the imaginary number $i$ in $e^\Phi$ turns type IIB superstring theory into the so-called type IIB${}^*$ string theory~\cite{Hull:1998vg}, which introduces potential instabilities. Note that such a timelike T-duality transformation should be more accurately treated as an analytic continuation instead of an equivalence relation. The factor ``$i$'' in the dilaton is typically relocated to be in the RR potentials, which in supergravity changes the signs of the RR kinetic terms~\cite{Hull:1998vg}. Here, we stick to the prescription~\eqref{eq:mm1tcb} only to make the relations between different corners of string theory manifest. In the $\omega \rightarrow \infty$ limit of the IIB${}^*$ string theory, we are led to the so-called Matrix (-1)-brane Theory (M(-1)T)~\cite{Blair:2023noj, Gomis:2023eav}. The fundamental degrees of freedom are now captured by the D-instantons (\emph{i.e.}~D(-1)-brane), whose dynamics is described by the IKKT matrix theory~\cite{Ishibashi:1996xs}. 

We now consider the fundamental string in M(-1)T. Even though its fundamental role is now replaced with the D-instanton, the M(-1)T string still encodes important information. The M(-1)T string~\cite{Gomis:2023eav, Blair:2025nno} shares common features with tensionless string theory, which in the Polyakov formulation is described by the action~\cite{Lindstrom:1990qb, Isberg:1993av}
\be \label{eq:tlsa}
    S^{}_\text{\scalebox{0.8}{P}} = \frac{T}{2} \int \dd^2 \sigma \, \gamma \, \gamma^\alpha{}^{}_0 \, \gamma^\beta{}^{}_0 \, \p^{}_\alpha X^\mu \, \p^{}_\beta X^\nu \, e^{}_{\mu\nu}\,.
\ee
Here, $\gamma_{\alpha}{}^0$ and $\gamma_\alpha{}^1$ are respectively the temporal and spatial the vielbein field on the worldsheet, which describe a two-dimensional Carrollian geometry~\cite{Bagchi:2015nca}. Moreover, we have defined $\gamma \equiv \det (\gamma^{}_\alpha{}^0 \,\, \gamma^{}_\alpha{}^1)$ and the inverse vielbein fields via
\be
    \gamma^\alpha{}^{}_0 \, \gamma^{}_\alpha{}^0 = \gamma^\alpha{}^{}_1 \, \gamma^{}_\alpha{}^1 = 1\,,
        \qquad%
    \gamma^\alpha{}^{}_0 \, \gamma_\alpha{}^1 = \gamma^\alpha{}^{}_1 \, \gamma^{}_\alpha{}^0 = 0\,,
        \qquad%
    \gamma^{}_\alpha{}^0 \, \gamma^\beta{}^{}_0 + \gamma^{}_\alpha{}^1 \, \gamma^\beta{}^{}_1 = \delta_\alpha^\beta\,.
\ee
The action~\eqref{eq:tlsa} is invariant under the worldsheet Carrollian boost transformation, which acts on the worldsheet vielbein fields as
\be
    \delta \gamma_\alpha{}^0 = \beta \, \gamma_\alpha^1\,,
        \qquad%
    \delta \gamma_\alpha{}^1 = 0\,,
        \qquad%
    \delta \gamma^\alpha{}^{}_0 = 0\,,
        \qquad%
    \delta \gamma^\alpha{}^{}_1 = - \beta \, \gamma^\alpha{}^{}_0\,.
\ee
The worldsheet topologies are the Riemann nodal spheres~\cite{Geyer:2015bja}. 
Recast in the phase space formulation, the tensionless string action becomes
\be
    S^{}_\text{p.s.} = \int \dd^2\sigma \, \Bigl( P^{}_\mu \, \p^{}_\tau X^\mu + \tfrac{1}{2} \, \chi \, P^{}_\mu \, P^{}_\nu \, e^{\mu\nu} - \rho \, P^{}_\mu \, \p^{}_\sigma X^\mu \Bigr)\,,
\ee
with $\sigma^\alpha = (\tau\,, \sigma)$\,. Under the gauge choice $\rho=1$ we are led to the chiral string action~\cite{Casali:2016atr},
\be \label{eq:boch}
    S^{}_\text{chiral} = \int \dd^2\sigma \, P^{}_\mu \, \bar{\p} X^\mu \,,
        \qquad%
    \CH = P^{}_\mu \, P^{}_\nu \, e^{\mu\nu} \sim 0\,,
\ee
where $\bar{\p} = \p^{}_\tau - \p^{}_\sigma$\,. We are now led to a chiral theory that is a curved $\beta\gamma$-system. This is a bosonic ambitwistor string theory, which does not contain any string vibration and the moduli space for string loops is localized on solutions to scattering equations. For appropriate vertex operators, this leads to the CHY formalism for particle scatterings~\cite{Cachazo:2013hca}. The associated string amplitudes naturally encode certain target space dynamics. 

An intriguing development that reveals how target space gravitational field equations arise from the ambitwistor string worldsheet theory was given in~\cite{Adamo:2014wea} (see also~\cite{Berkovits:2018jvm}), which requires extending the bosonic action~\eqref{eq:boch} to be supersymmetric.\,\footnote{It is also known that the bosonic ambitwistor string leads to wrong gravitational equations of motion with higher derivatives~\cite{Berkovits:2018jvm}.}
In flat spacetime, the Ramond-Neveu-Schwarz (RNS) formulation of the tensionless superstring is given by
\be \label{eq:schiral}
    S^{}_\text{chiral} = \int \Bigl( P^{}_\mu \, \bar{\p} X^\mu + \bar{\psi}^{}_\mu \, \bar{\p} \psi^\mu \Bigr)\,,
\ee
with $\psi^\mu$ a complex fermion. This chiral action is supplemented with the constraints,
\be
    \CG \equiv \psi^\mu \, P^{}_\mu \sim 0\,,
        \qquad%
    \bar{\CG} \equiv \bar{\psi}^{}_\mu \, P^{}_\nu \, \eta^{\mu\nu} \sim 0\,,
        \qquad%
    \CH \equiv P^{}_\mu \, P^{}_\nu \, \eta^{\mu\nu}\sim 0\,.
\ee
This worldsheet theory is invariant under the supersymmetric transformations
\be
    \delta X^\mu = \epsilon \, \eta^{\mu\nu} \, \bar{\psi}^{}_\nu + \bar{\epsilon} \, \psi^\mu,
        \qquad%
    \delta \psi^\mu = - \epsilon \, \eta^{\mu\nu} \, P^{}_\nu - \bar{\epsilon} \, P^\mu\,.
\ee
Note that $\CG$ and $\bar{\CG}$ are supercurrents that are constrained to vanish. The operator product expansion (OPE) between these supercurrents give
\be \label{eq:ggh}
    : \CG (z) \, \bar{\CG} (w) : \, \sim \frac{\CH (w)}{z - w}\,,
\ee
implying that the constraints form a closed algebra at the quantum level. 

The generalization to the curved $\beta\gamma$-system is much trickier, where various extensions of the supercurrents are required in order for all the fields to transform correctly under both target space and worldsheet diffeomorphisms at the quantum level. These improved supercurrents behave anomalously: the OPE $: \! \CG(z) \, \bar{\CG}(w) \! :$ contains higher-order poles. The closedness of the constraint algebra then requires that the coefficient of any higher-order pole vanish, such that Eq.~\eqref{eq:ggh} continues to hold for the improved supercurrents and quantum corrected Hamiltonian $\CH$. The anomaly-free conditions are given in~\cite{Adamo:2014wea}, where a particular construction of the supercurrents is adopted, to coincide with the NSNS sector of ten-dimensional supergravity field equations. However, for our application to M(-1)T, a different set of symmetries including the emergent dilatation are involved, which implies that the construction of the supercurrents would differ. While the metric dependence in the equations of motion remains unchanged, how the dilaton and Kalb-Ramond field enter will be altered.  
This distinction can be made manifest by 
analyzing the effective gravity in the target space. Consider the following IIB supergravity action, focusing on the sector containing the metric $G_{\mu\nu}$\,, dilaton $\Phi$\,, and RR 0-form field $C^{(0)}$\,:
\begin{align} \label{eq:siibg}
	S^{}_{\text{\scalebox{0.8}{IIB}}} &= \frac{1}{2 \, \kappa^2} \int \dd^{10} x \, \sqrt{-G} \, \biggl[ e^{-2\Phi} \Bigl( R + 4 \, \p_\mu \Phi \, \p^\mu \Phi \Bigr) -\tfrac{1}{2} \, \p_\mu C^{(0)} \, \p^\mu C^{(0)} \biggr].
\end{align}
Using the ansatz~\eqref{eq:mm1tcb} and taking the $\omega \rightarrow \infty$ limit, we find 
\be \label{eq:iibsdl}
    S^{}_\text{\scalebox{0.8}{IIB}} \rightarrow - \frac{1}{2 \, \kappa^2} \int \dd^{10} x \, e \, e^{-2\varphi} \, \Bigl( r + \tfrac{9}{2} \, \p_\mu \varphi \, \p^\mu \varphi \Bigr),
\ee
with $e = \det e^{}_\mu{}^a$ and $r$ the Ricci scalar for $e^{}_{\mu\nu}$\,. This theory is Lorentz invariant. Moreover, it is also invariant an emergent dilatation symmetry acting non-trivially as $e^{}_{\mu\nu} \rightarrow \Omega^{-1} \, e^{}_{\mu\nu}$ and $e^\varphi \rightarrow \Omega^{-2} \, e^\varphi$, up to a boundary term
$- (9 / 2\kappa^2) \int \dd^{10} x \, \p_\mu \bigl( e \, e^{-2\varphi} \, e^{\mu\nu} \, \p^{}_\nu \ln \Omega \bigr)$\,,
with $e^{\mu\nu}$ the inverse of $e^{}_{\mu\nu}$\,. In the original formalism~\cite{Hull:1998vg}, the ``wrong'' sign in Eq.~\eqref{eq:iibsdl} would be flipped, accompanied by the analytic continuation $c^{(q)} \rightarrow - i \, c^{(q)}$. Interestingly, the potentially problematic kinetic term associated with $c^{(0)}$ drops out of the resulting action~\eqref{eq:iibsdl} after the decoupling limit is performed.  
Note that T-duals of this Lorentzian supergravity give rise to various non-Lorentzian structures: a timelike T-duality transformation (or, an analytic continuation) maps the theory to Eq.~\eqref{eq:expsiia2}, which is Galilean, while a spatial T-duality transformation maps it to a Carroll-like theory~\cite{Blair:2023noj, Gomis:2023eav, Blair:2025nno}.  

We now comment on how to apply the above techniques to the M0T string in order to extract the target space non-Lorentzian gravitational equations of motion. As an initial step, we focus on flat spacetime. The bosonic part of the M0T string can be obtained by formally T-dualizing the tensionless string action~\eqref{eq:tlsa}, which gives~\cite{Gomis:2023eav}\,\footnote{See also~\cite{Albrychiewicz:2023ngk} in the context of tropological sigma models.}
\be \label{eq:m0ts}
    S = \frac{T}{2} \int \dd^2 \sigma \, \sqrt{-\gamma} \, \Bigl[ \bigl( \gamma^\alpha{}^{}_1 \, \p^{}_\alpha X^0 \bigr)^2 + \gamma^\alpha{}^{}_0 \, \gamma^\beta{}^{}_0 \, \p^{}_\alpha X^i \, \p^{}_\beta X^i + \lambda \, \gamma^\alpha{}^{}_0 \, \p^{}_\alpha X^0 \Bigr]\,.
\ee
Note that integrating out the worldsheet zweibein $\gamma_\alpha{}^a$ gives rise to the Nambu-Goto formulation that coincides with the leading order term in Eq.~\eqref{eq:fos}~\cite{Batlle:2016iel}. For convenience we set $T=2$ below. 
In the first-order formulation, we choose a gauge as in Eq.~\eqref{eq:boch} with $\rho = 1$\,, we find the chiral action~\cite{Gomis:2023eav},
\be \label{eq:bochm0t}
    S^{}_\text{chiral} = \int \dd^2\sigma \, P^{}_\mu \, \bar{\p} X^\mu \,,
        \qquad%
    \CH = P^{}_i \, P^{}_i - \p X^0 \, \p X^0 \sim 0\,,
\ee
which is invariant under the Galilei boost $\delta_\text{\scalebox{0.8}{G}} X^i = v^i \, X^0$ and $\delta_\text{\scalebox{0.8}{G}} P^{}_0 = - v^i \, P^{}_i$\,. 
Its supersymmetric generalization is in form the same as Eq.~\eqref{eq:schiral}, but with different constraints,
\be
    \CG = \psi^i \, P^{}_i + \psi^0 \, \p X^0 \sim 0\,,
        \qquad%
    \bar{\CG} = \bar{\psi}^{}_i \, P^{}_i - \bar{\psi}^{}_0 \, \p X^0 \sim 0\,,
        \qquad%
    \CH = P^{}_i \, P^{}_i - \p X^0 \, \p X^0 \sim 0\,.
\ee
The associated supersymetric transformations are now
\begin{subequations}
\begin{align}
    \delta P^{}_0 &= - \p \bigl( \epsilon \, \bar{\psi}^{}_0 \bigr) + \p \bigl( \bar{\epsilon} \, \psi^0 \bigr)\,,
        &%
    \delta \psi^0 &= \epsilon \, \p X^0,
        &%
    \delta \bar{\psi}^{}_0 &= - \bar{\epsilon} \, \p X^0, \\[4pt]
    \delta X^i &= \epsilon \, \delta^{ij} \, \bar{\psi}^{}_j + \bar{\epsilon} \, \psi^i,
        &%
    \delta \psi^i &= - \epsilon \, \delta^{ij} P^{}_j\,,
        &%
    \delta \bar{\psi}^{}_i &= - \bar{\epsilon} \, P^{}_i\,.
\end{align}
\end{subequations}
Note that the closedness condition~\eqref{eq:ggh} is satisfied. Generalizing this construction to include curved background fields should lead to anomaly-free conditions that correspond to the equations of motion associated with the non-Lorentzian supergravity~\eqref{eq:expsiia2}. How the details work out is still highly non-trivial, which we leave for future studies. 

\subsection{Brane Dynamics on Curved Non-Lorentzian Geometry} \label{sec:bdcnlg}

After describing the framework of M0T and its low-energy supergravity, we now look into the dynamics of the non-Lorentzian supergravity~\eqref{eq:expsiia2}. We will consider the backreaction of string and brane objects in the M0T supergravity and construct curved non-Lorentzian geometry solutions, as well as the associated scattering processes. 

\vspace{3mm}

\noindent $\bullet$~\textbf{D-Particle Backreaction.} We now consider the D-brane source in M0T described by~\eqref{eq:d0a} and focus on a static bound state, which has the charge $\int \dd t \, \p_t X^\mu (t) \, c^{}_\mu$\,, with $X^\mu (t) = (t\,,\mathbf{0})$\,. Varying the total action that couples $N$ static D-particle source to the non-Lorentzian supergravity~\eqref{eq:expsiia2} with respect to the RR potential $c^{}_\mu$\,, and then projecting by the temporal vielbein $\tau^{}_\mu$\,, gives rise to the following equation of motion:
\be \label{eq:teom}
	t^{}_{ij} \, t^{}_{ij} \, \propto \, N \, e^\varphi \, \delta^{(9)}\bigl( x^i \bigr)\,,
        \qquad%
    t^{}_{ij} \equiv e^\mu{}^{}_i \, e^\nu{}^{}_j \, \p^{}_{[\mu} \tau^{}_{\nu]}\,.
\ee
We have used Eq.~\eqref{eq:eomc0} to obtain the above expression. 
This equation of motion implies that the D-particles source the torsion $t_{ij}$\,. 
In general, the twistless torsion condition $t_{ij} = 0$ is essential for preserving the foliation structure, and its violation has to be treated with care as this may introduce pathological behaviors~\cite{Hartong:2015zia}. Moreover, it is unclear how the nonlinear equation of motion~\eqref{eq:teom} can be solved. 

Intuitions about how such a backreaction should be dealt with comes from non-relativistic string theory, where a codimension-two foliation is introduced. An extensive review of the relevant ingredients from non-relativistic string theory is given in Appendix~\ref{app:pnrst}. From the worldsheet perspective of non-relativistic string theory, the violation of the twistless torsion condition by quantum corrections generates a deformation towards the full relativistic string theory~\cite{Gomis:2019zyu, Yan:2021lbe}. Similarly, in M0T, the D-particle sourcing should also be considered in the full type IIA superstring theory rather than the limiting case. This requires us to turn back on a finite $\omega$ in the M0T limiting prescription, and treat it as a spacetime-dependent background field. At the linearized order of the fluctuations around flat spacetime (\emph{i.e.}~$\tau^{}_\mu = \delta_\mu^0$\,, $e^{}_\mu{}^i = \delta_\mu^i$\,, $e^\varphi = g^{}_\text{s}$\,, $b^{(2)} = 0$\,, and $c^{(q)} = 0$), this leads us to the Poisson equation,
\be \label{eq:pe}
	\p^{}_i \, \p^{}_i \bigl( \omega^{-2} \bigr) \, \propto \, g^{}_\text{s} \, N \, \delta^{(9)} (x^i)\,,
\ee
which is solved by $\omega^{-2} \sim (\ell / r)^{7}$, with $\ell^7 \sim g^{}_\text{s} \, N$. Plugging this solution back into Eq.~\eqref{eq:m0tcb} with the flat conditions reproduces the standard D-particle geometry Eq.~\eqref{eq:nhg0}. Even though we are focusing on the bosonic sector, it is the target space supersymmetry and the analysis of the Killing spinor equation that allow one to easily identify the Poisson equation~\eqref{eq:pe} with a simple form.  

It is interesting to compare the above scenario with its analog in the dual frame of non-relativistic string theory. In this dual frame, the non-Lorentzian supergravity as the low-energy description does not contain any propagating gravitational mode, either. Instead, it describes instantaneous Newton-like force between winding strings. In the zero-winding sector described by the so-called `string Newton-Cartan gravity' underlying non-relativistic string theory~\cite{Andringa:2012uz, Bergshoeff:2019pij, Bidussi:2021ujm}, there is \emph{no} asymptotic string state~\cite{Gomis:2000bd, Danielsson:2000gi}. If one backreacts the fundamental string in string Newton-Cartan gravity, then the theory is deformed back to type IIB superstring theory. 
This $\omega$-deformation acquires an interpretation as the $T\bar{T}$ deformation in the dual description of non-relativistic string theory (\emph{i.e.}~matrix string theory in its second-quantized form)~\cite{Blair:2020ops, Blair:2024aqz}. 
Therefore, the computation of the instantaneous gravitational interactions between the non-relativistic strings with nonzero windings is beyond the string Newton-Cartan gravity. Instead, a string amplitude computation is required~\cite{Gomis:2000bd, Yan:2021hte}. See Appendix~\ref{app:pnrst} for further discussions. 

Dualizing non-relativistic string theory back to M0T, the associated non-Lorentzian supergravity now describes the instantaneous Newtonian force between D-particles, which can be computed by analyzing string amplitudes between D-particles or directly by solving the BFSS matrix theory. As we have seen, backreacting the D-particles deform the theory back to type IIA supergravity. This might suggests that there lacks interesting dynamics in non-Lorentzian supergravity, as one cannot source the gravitational force with the fundamental degree of freedom, may it be the fundamental string in non-relativistic string theory or the D-particle in M0T. On the contrary, it has been noted in~\cite{Lambert:2024uue, Lambert:2024yjk, Lambert:2024ncn, Blair:2024aqz, Harmark:2025ikv, Blair:2025ewa} that interesting curved non-Lorentzian geometries can be constructed from backreacting extended objects other than the fundamental degrees of freedom. We will examine such sources in non-Lorentzian supergravity in the next step. 

\vspace{3mm}

\noindent $\bullet$~\textbf{Non-Lorentzian F1-string geometry.} We start with backreacting the M0T string, which we have discussed around Eq.~\eqref{eq:m0ts}. In arbitrary curved background fields and focusing on the bosonic sector, its Nambu-Goto formulation is given by~\cite{Batlle:2016iel}
\be \label{eq:m0ts2}
    S^{\text{\scalebox{0.8}{M0T}}}_{\text{\scalebox{0.8}{F1}}} = - \frac{1}{\ell^2_\text{s}} \int \dd^2 \sigma \, \sqrt{\epsilon^{\alpha\beta} \, \epsilon^{\gamma\delta} \, \tau^{}_\alpha \, \tau^{}_\gamma \, e^{}_{\beta\delta}} - \frac{1}{\ell^2_\text{s}} \int b^{(2)}.
\ee
Coupling $N$ M0T strings to the non-Lorentzian supergravity~\eqref{eq:expsiia2} form the total action $S^{}_\text{NL} + N S^{\text{\scalebox{0.8}{M0T}}}_{\text{\scalebox{0.8}{F1}}}$\,. We consider a static string source. 
Instead of directly solving the associated equations of motion, it is easier to infer the solution by taking an M0T limit of the following string soliton geometry~\cite{Dabholkar:1990yf} from solving the IIA supergravity equations of motion:
\begin{align} \label{eq:ss}
    \dd s^2 &= \frac{- \dd \mathbf{x}_0^2 + \dd \mathbf{x}^2_1}{H_\text{\scalebox{0.8}{F1}}} + \dd R^2 + R^2 \, \dd\Omega^2_7\,,
        \qquad%
    B^{(2)} = - \frac{\dd \mathbf{x}^0 \wedge \dd \mathbf{x}^1}{H_\text{\scalebox{0.8}{F1}}},
        \qquad%
    e^\Phi = \frac{G^{}_\text{s}}{\sqrt{H_\text{\scalebox{0.8}{F1}}}}\,,
\end{align}
with $i = 2\,,\,\cdots, \,9$ and $H_\text{\scalebox{0.8}{F1}} = 1 + L^6 / R^6$, where $L^6 \sim N \, G^2_\text{s}$ and $R^2 = \mathbf{x}^i \, \mathbf{x}^i$. In order to take an M0T limit, an extra critical RR 0-form potential $C^{(1)}$ has to be included, which could potentially spoil the solution. However, since F1-D0 bound state is $\frac{1}{4}$-BPS, this $C^{(1)}$ has to be pure gauge. Applying the BPS decoupling limit~\eqref{eq:m0tlp} zooming in on the D0-brane asymptotically at $R \rightarrow \infty$\,, we find 
\begin{align}
    \dd s^2_{\infty} &= \dd \mathbf{x}^\mu \, \dd \mathbf{x}^{}_\mu = - \omega \, \dd t^2 + \omega^{-1} \, \bigl( \dd x^2_1 + \dd x^i \, \dd x^i \bigr)\,, \\[4pt]
    B^{(2)}_\infty &= - \dd \mathbf{x}^0 \wedge \dd \mathbf{x}^1 = - \dd x^0 \wedge \dd x^1, 
        \qquad%
    e^{\Phi_\infty} = G^{}_\text{s} = \omega^{-3/2} \, g^{}_\text{s}\,.
\end{align}
It can be inferred that, consistently, $C^{(1)}_\infty = \omega^2 \, g^{-1}_\text{s} \, \dd t$\,. Therefore,
\be \label{eq:cxlx}
    \mathbf{x}^0 = \omega^{1/2} \, t\,,
        \qquad%
    \mathbf{x}^1 = \frac{x^1}{\omega^{1/2}}\,,
        \qquad%
    \mathbf{x}^i = \frac{x^i}{\omega^{1/2}}\,,
        \qquad%
    G^{}_\text{s} = \frac{g^{}_\text{s}}{\omega^{3/2}}\,,
\ee
which prescribes the M0T limit~\eqref{eq:m0tlp} in the asymptotic infinity. 
Plugging Eq.~\eqref{eq:cxlx} into the string soliton geometry~\eqref{eq:ss}, we are led to the M0T limiting prescription~\eqref{eq:m0tcb}. In the $\omega \rightarrow \infty$ limit, this gives rise to the following non-Lorentzian geometry:
\begin{subequations} \label{eq:tef}
\begin{align} 
    \tau^{}_\mu \, \dd x^\mu &= \frac{\dd t}{\sqrt{H_\text{\scalebox{0.8}{F1}}}},
        \qquad%
    e^{}_{\mu\nu} \, \dd x^\mu \, \dd x^\nu = \frac{\dd x_1^2}{H_\text{\scalebox{0.8}{F1}}} + \dd r^2 + r^2 \, \dd \Omega_7^2\,, 
     \\[4pt]
    e^\varphi &= \frac{g^{}_\text{s}}{\sqrt{H_\text{\scalebox{0.8}{F1}}}}\,,
        \qquad%
    b^{(2)} = - \frac{\dd t \wedge \dd x^1}{H_\text{\scalebox{0.8}{F1}}},
        \qquad%
    H_\text{\scalebox{0.8}{F1}} = 1 + \frac{\ell^6}{r^6}\,,
\end{align}
\end{subequations}
where $\ell^6 = g^2_\text{s} \, \ell_0^6$\,, $r^2 = x^2_2 + \cdots + x^2_9$\,, and $e^{}_{\mu\nu} = e^{}_\mu{}^1 \, e^{}_\nu{}^1 + e^{}_\mu{}^i \, e^{}_\nu{}^i$\,. We further infer that 
\be \label{eq:crro}
    C^{(1)} = \omega^2 \, e^{-\varphi} \, \tau^{}_\mu \, \dd x^\mu = \omega^2 \, g^{-1}_\text{s} \, \dd t 
\ee
has zero flux, which does not spoil Eq.~\eqref{eq:tef} as a solution to the non-Lorentzian supergravity~\eqref{eq:expsiia2}. 
It is straightforward to check that Eq.~\eqref{eq:tef} is a solution to the equations of motion associated with the variations in Appendix~\ref{app:eom}, which arise from varying the fields in the non-Lorentzian supergravity action~\eqref{eq:expsiia2}, but now with the F1 source included. 

We now consider a probe D-particle moving on the string soliton background~\eqref{eq:tef}. The D-particle action on this M0T geometry is 
\be
    S^{\text{\scalebox{0.8}{M0T}}}_{\text{\scalebox{0.8}{D0}}} = \frac{1}{2 \, g^{}_\text{s} \, \ell_\text{s}} \int \dd t \, \frac{1}{\dot{X}^0} \ls \dot{X}^2_1 + H_\text{\scalebox{0.8}{F1}} \, \bigl( \dot{X}^2_2 + \cdots + \dot{X}^2_9 \bigr) \rs. 
\ee 
Take the D-particle trajectory to be $X^\mu = (t, 0, \cdots, 0, v \, t)$ such that it moves orthogonally to the background F1-string in the transverse directions, we find the effective action 
\be \label{eq:smtdm2}
    S^{\text{\scalebox{0.8}{M0T}}}_{\text{\scalebox{0.8}{D0}}} = \frac{1}{2 \, g^{}_\text{s} \, \ell_\text{s}} \int \dd t \ls 1 + \frac{g^2_\text{s}}{(r/\ell_0)^6} \rs v^2\,,
\ee
which describes a non-relativistic particle in an attractive potential $\sim v^2 / r^6$.  
In contrast, when the D-particle moves in parallel to the F1-string, the effective action reduces to the kinetic energy of a free D-particle. 

It is also interesting to consider a static probe M0T string~\eqref{eq:m0ts2} on the M0T string soliton background~\eqref{eq:tef},
\begin{align} \label{eq:m0tsf}
    S^{\text{\scalebox{0.8}{M0T}}}_{\text{\scalebox{0.8}{F1}}} = & - \frac{1}{\ell^2_\text{s}} \int \dd^2 \sigma \, \frac{1}{H_\text{\scalebox{0.8}{F1}}}\sqrt{\epsilon^{\alpha\beta} \, \epsilon^{\gamma\delta} \, \p^{}_\alpha X^0 \, \p^{}_\gamma X^0 \, \Bigl( \p_\beta X^1 \, \p_\delta X^1 + H_\text{\scalebox{0.8}{F1}} \, \p_\beta X^i \, \p_\delta X^i \Bigr)} \notag\\[4pt]
	& - \frac{1}{\ell^2_\text{s}} \int \frac{\dd X^0 \wedge \dd X^1}{H_\text{\scalebox{0.8}{F1}}},
		\qquad%
	i = 2\,,\,\cdots,\,9\,.
\end{align}
In the case where the probe M0T string is positioned in parallel with the background M0T strings, \emph{i.e.}~$X^\mu = (\sigma^\alpha, 0\,,\,\cdots,0)$\,, the potential exactly vanishes due to the critical Kalb-Ramond field. In contrast, when the probe M0T string is positioned orthogonally to the background M0T strings, \emph{i.e.}~$X^\mu = (\sigma^0\,, \,0\,, \cdots, 0\,, \, \sigma^1)$\,, we find
\be
	S^{\text{\scalebox{0.8}{M0T}}}_{\text{\scalebox{0.8}{F1}}} = - \frac{1}{\ell^2_\text{s}} \int \dd^2 \sigma \ls 1 + \frac{g^2_\text{s}}{(r/\ell_0)^6} \rs^{\!-1/2} = \frac{1}{\ell^2_\text{s}} \int \dd^2 \sigma \ls - 1 + \frac{g^2_\text{s}}{2 \, (r/\ell_0)^6} + O\bigl(r^{-12}\bigr) \rs.
\ee
At the leading order, there is an attractive potential $\sim r^{-6}$ between the orthogonal strings.  

\vspace{3mm}

\noindent $\bullet$~\textbf{Non-Lorentzian  D4-brane geometry.}
Since D0-D4 also forms a $\frac{1}{4}$-BPS configuration, backreacting D4-branes should also give rise to the following curved geometry in M0T supergravity~\cite{Lambert:2024yjk, Blair:2024aqz}:
\begin{subequations} \label{eq:m0td4}
\begin{align}
    \tau^{}_\mu \, \dd x^\mu &= \frac{\dd t}{H_4^{1/4}}\,,
        &%
    e^{}_{\mu\nu} \, \dd x^\mu \, \dd x^\nu &= \frac{\dd x_1^2 + \cdots + \dd x_4^2}{\sqrt{H^{}_4}} + \sqrt{H^{}_4} \, \bigl( \dd r^2 + r^2 \, \dd \Omega_4^2 \bigr)\,, \\[4pt]
    e^\varphi &= \frac{g^{}_\text{s}}{H_4^{1/4}}\,,
        &%
    c^{(5)} &= \frac{1}{g^{}_\text{s} \, H^{}_4} \, \dd t \wedge \cdots \wedge \dd x^4,
        \qquad%
    H^{}_4 = 1 + \frac{\ell^3}{r^3}\,.
\end{align}
\end{subequations}
with $\ell^3 = g^{}_\text{s} \, \ell_0^3$ and $r^2 = x_5^2 + \cdots + x_9^2$\,. Note that the critical RR one-form in the limiting prescription is inferred to take the same form as in Eq.~\eqref{eq:crro}, which is a pure gauge that does not spoil the supergravity solution. 
This is the geometry that arises from backreacting D4-branes in M0T supergravity, which solves the M0T supergravity equations of motion. 

We now consider a probe D-particle on this non-Lorentzian D4-brane geometry. On the M0T geometry~\eqref{eq:m0td4} the D-particle action~\eqref{eq:d0a} takes the form
\be
    S^{\text{\scalebox{0.8}{M0T}}}_{\text{\scalebox{0.8}{D0}}} = \frac{1}{2 \, g^{}_\text{s} \, \ell_\text{s}} \int \dd t \, \frac{1}{\dot{X}^0} \ls \dot{X}^2_1 + \cdots + \dot{X}^2_4 + H^{}_4 \, \bigl( \dot{X}^2_5 + \cdots + \dot{X}^2_9 \bigr) \rs,
\ee 
Take the D-particle trajectory to be $X^\mu = (t, 0, \cdots, 0, v \, t)$\,, \emph{i.e.}~the D-particle moves orthogonally to the background D4-brane in the transverse directions, we find the effective action 
\be \label{eq:smtdm}
    S^{\text{\scalebox{0.8}{M0T}}}_{\text{\scalebox{0.8}{D0}}} = \frac{1}{2 \, g^{}_\text{s} \, \ell_\text{s}} \int \dd t \ls 1 + \frac{g^{}_\text{s}}{(r / \ell_0)^3} \rs v^2\,.
\ee
Just like in the case of an M0T string background, when the D-particle moves in parallel to the D4-brane the effective action becomes trivial. There is an attractive potential $\sim v^2 / r^3$\,. 

It is expected that this result matches the amplitude from scattering a D-particle on a D4-brane, with the D-particle moving at a small velocity $v$ orthogonal to the D4-brane~\cite{Douglas:1996yp}. In the eikonal regime and at the one-loop order, the full scattering amplitude in type IIA superstring theory is described by the phase shift $e^{i \delta (b, v)}$, with $b$ the impact parameter and~\cite{Lifschytz:1996iq, Douglas:1996yp}
\be
    \delta (b, v) = \int_0^\infty \frac{\dd s}{s} \, e^{- s \, b^2} \, \tan \frac{s \, v}{2}\,.
\ee
For the straight line trajectory of the D-particle, we have $(r/\ell_0)^2 = b^2 + v^2 \, t^2$. Together with the tree-level term, \emph{i.e.}~without interaction, the associated effective action is
\begin{align}
\begin{split}
    S &= \ell^{-1}_\text{s} \int \dd t \lr \frac{v^2}{2 \, g^{}_\text{s}} + v \! \int_0^\infty \frac{\dd s}{\sqrt{\pi s}} \, e^{-s \, (r/\ell_0)^2} \, \tan \frac{s \, v}{2} \rr \\[4pt]
    &= \frac{1}{2 \, g^{}_\text{s} \, \ell^{}_\text{s}} \int \dd t \ls 1 + \frac{g^{}_\text{s}}{(r / \ell_0)^3} \rs v^2 + O(v^4 / r^7)\,,
\end{split}
\end{align}
where we have recovered the dependence on $\ell_\text{s}$ and the leading order term exactly matches Eq.~\eqref{eq:smtdm}. When the D-particle moves in parallel to a D4-brane, the amplitude trivializes.  

Next, we consider a probe string with $X^0 = \sigma^0$. On the D4-brane background~\eqref{eq:m0td4}, the probe M0T string action~\eqref{eq:m0ts2} becomes
\begin{align} \label{eq:m0tsfdf}
    	S^{\text{\scalebox{0.8}{M0T}}}_{\text{\scalebox{0.8}{F1}}} = & - \frac{1}{\ell^2_\text{s}} \int \dd^2 \sigma \, \sqrt{\frac{X'_1 \, X'_1 + \cdots + X'_4 \, X'_4}{H^{}_4} + X'_5 \, X'_5 + \cdots + X'_9 \, X'_9}\,.
\end{align}
While a static M0T string positioned orthogonally to the M0T D4-brane background has a trivial potential, the one parallel to the D4-brane, say, with $X^1 = \sigma^1$, develops a non-trivial effective action,
\be
	S^{\text{\scalebox{0.8}{M0T}}}_{\text{\scalebox{0.8}{F1}}} = - \frac{1}{\ell^2_\text{s}} \int \dd^2 \sigma \ls 1 + \frac{g^{}_\text{s}}{(r/\ell_0)^3} \rs^{\!-1/2} = \frac{1}{\ell^2_\text{s}} \int \dd^2 \sigma \ls - 1 + \frac{g^{}_\text{s}}{2 \, (r/\ell_0)^3} + O\bigl(r^{-6}\bigr) \rs,
\ee
\emph{i.e.}~an effective attractive potential $\sim r^{-3}$ at large $r$\,. 
The same effective potential can be reproduced by considering a probe D4-brane on the M0T F-string background~\eqref{eq:tef}, where smearing of the background geometry and matching the F-string flux is required. 

\vspace{3mm}

\noindent $\bullet$~\textbf{Other branes.} In the above examples of curved non-Lorentzian geometry, it is important that the harmonic function retains a non-trivial structure. For general D$p$-branes with an even $p$\,, the harmonic function is $H = 1 + (L / R)^{7-p}$, with $L^{7-p} \sim N \, G^{}_\text{s}$\,. Applying the limiting prescription~\eqref{eq:cxlx} leads to $H = 1 +  \omega^{(4-p)/2} \, (\ell / r)^{7-p}$. Only when $p = 4$, in which case the D0-D4 bound state is $\frac{1}{4}$-BPS, the harmonic function $H$ retains a non-trivial form with $H \rightarrow 1 + (\ell / r)^3$. As a result, an asymptotic M0T limit does not give rise to a bulk M0T limit that leads to a curved non-Lorentzian geometry if $p \neq 4$. Moreover, for the NS5-brane geometry, the harmonic function is reparametrized to be $H \rightarrow 1 + \omega \, (\ell/r)^2$, also spoiling the M0T structure. 

This obstruction of constructing a curved brane geometry in M0T supergravity when the brane does not form a BPS state with the D-particle can also be understood in a more dynamical way. We take D2-brane as an example. The static source term in the equation of motion arises from varying with respect to the electric part of the RR 3-form $c^{}_{0ij}$\,. In the absence of any background string, \emph{i.e.}~with zero $B$-field $b^{(2)} = 0$\,, Eq.~\eqref{eq:eomc0ij} implies that the the D2-brane density sources the torsion, 
\be
     t^{}_{k\ell} \, \bigl( \dd c^{(3)} \bigr)_{ijk\ell} \,\propto\, N \, \epsilon^{}_{ij} \, \delta^{(7)}(x)\,,
\ee
Just like the backreaction of the D-particles, this implies that the static D2-branes induce instability in non-Lorentzian supergravity via sourcing the torsion, which generates a flow towards the IIA theory. As a result, we recover the standard D2-brane geometry that is Lorentzian. Moreover, one may also consider the backreaction of orthogonal brane states that form a BPS object with the D-particles, which we leave it to future studies. 

\vspace{3mm}

We note that the discussions in this section are generalizable to other BPS decoupling limits zooming on different extended objects, in connection with more general matrix (gauge) theories. See \emph{e.g.}~\cite{Lambert:2024uue, Lambert:2024yjk, Lambert:2024ncn, Blair:2024aqz, Harmark:2025ikv, Blair:2025ewa}. 

\section{Back to Holography} \label{sec:bth}

In the previous section we discussed the BPS decoupling limit of type IIA superstring theory zooming in on a D-particle state, \emph{i.e.}~Matrix 0-brane Theory (M0T). The fundamental degrees of freedom are non-relativistic D-particles, whose dynamics is described by perturbative BFSS matrix theory. We also studied the low-energy effective supergravity description in M0T and argued that its dynamics should arise from analyzing the F-string. We showed that there exist non-trivial non-Lorentzian curved solutions from backreacting the F-strings and D4-branes. In contrast, the backreaction of D-particles in M0T supergravity goes beyond the non-Lorentzian framework and leads to the relativistic geometry~\eqref{eq:bnhg} in IIA supergravity, which contains a conformally AdS${}_2$ sector. This connects to the holographic description at large $N$. 

We now relate our analysis of M0T to the idea introduced in Section~\ref{sec:nlrdph}, where we started with the near-horizon D-particle geometry and then performed a large $\tilde{\omega}$ expansion. While the near-horizon limit corresponds to the BPS decoupling that leads to M0T, the $\tilde{\omega}$-expansion decouples the D-particles at the leading order, which avoids the M0T supergravity from being deformed towards the full IIA theory. On a flat background, this truncated M0T supergravity is equivalent to the null reduction of eleven-dimensional supergravity, which, according to Section~\ref{sec:mln}, holographically matches onto the non-Lorentzian regime that arises as the leading-order contribution of the large $\tilde{\omega}$ expansion of the bulk supergravity. 

In order for the holographic correspondence to be valid, we need a large $N$. On the other hand, in order for the $\tilde{\omega}$-expansion to be valid, we require that $N$ be `moderately' large, \emph{i.e.}~$1 \ll N \ll \tilde{\omega}^2$\,. This raises the question that how much one could trust the M0T analysis, which is only a good approximation at low-$N$ when BFSS matrix theory that is
perturbatively defined. Since the D-particle dynamics becomes subleading at moderately large $N$, the leading-order M0T supergravity contribution remains classical and valid. 
We therefore reach the following conclusion: at moderately large $N$, we are left with the same classical non-Lorentzian supergravity~\eqref{eq:nlg} on a flat background, on both the M0T and the bulk gravity side. The `holographic correspondence' between the truncations at $O(\tilde{\omega})$ on both sides therefore reduces to a trivial equivalence.

The above connection to the M0T supergravity allows us to complete the analysis in Section~\ref{sec:nlrdph} to include all the gauge potentials, which we detail in Section~\ref{sec:ddp}. We will then generalize the $\tilde{\omega}$-expansion to D$p$-brane and string soliton holography.

\subsection{Decouple the D-Particles} \label{sec:ddp}

Consider the M0T prescription~\eqref{eq:m0tcb}, which we now treat as a large $\omega$ expansion instead of the strict $\omega \rightarrow \infty$ limit. This is necessary when we consider the D-particle backreaction, as it reintroduces the $\omega$-deformation. Next, we introduce the $\tilde{\omega}$ rescaling in Eq.~\eqref{eq:gslsrs0}, which implies the prescription~\eqref{eq:ftoa1} with the extra rescaling of the string length $\ell_\text{s} \rightarrow \ell_\text{s} / \sqrt{\tilde{\omega}}$\,. We start with improving the discussion on the $\tilde{\omega}$-expansion in Section~\ref{sec:nlrdph}. 

Just like in Section~\ref{eq:nllst}, where we used the prescription~\eqref{eq:m0tlp} that absorbs the $\omega$-scaling of the string length in the original prescription~\eqref{eq:dlo} into the background fields, we first construct analogously an equivalent parametrization that absorbs the $\tilde{\omega}$-scaling of the string length into the background fields. This will make it easier for us to understand the geometric implications. A simple way to achieve this is by considering probe the F-string and D-branes described by the actions in Eq.~\eqref{eq:sba}. We therefore absorb the rescaling $\ell_\text{s} \rightarrow \ell_\text{s}/ \sqrt{\tilde{\omega}}$ into the background fields by replacing this rescaling with
\be \label{eq:rsgt}
	G^{}_{\alpha\beta} \rightarrow \tilde{\omega} \, G^{}_{\alpha\beta}\,,
		\qquad%
	\CT^{(q)} \rightarrow \tilde{\omega}^{\frac{q}{2}} \, \CT^{(q)},
\ee
for any $q$-form $\CT^{(q)}$. 
Combined with the prescription~\eqref{eq:ftoa1}, we find 
\begin{align} \label{eq:ftoa}
	\dd s^2 &= - \tilde{\omega}^2 \, \bigl( \tau^{}_\mu \, \dd x^\mu \bigr)^2 + e^{}_\mu{}^i \, e^{}_\nu{}^i \, \dd x^\mu \, \dd x^\nu,
        \qquad%
    e^\Phi = e^\varphi,
		\qquad%
	C^{(1)} = \frac{\tilde{\omega}}{e^{\varphi}} \, \tau^{}_\mu \, \dd x^\mu + \frac{c^{(1)}}{\tilde{\omega}}\,,
\end{align}
with $\langle e^\varphi \rangle = \tilde{g}^{}_\text{s} \, \tilde{h}^{3/4}$. Now, the string length $\ell_\text{s}$ is kept unchanged. Here, $\varphi$ and $c^{(1)}$ will respectively be the dilaton and RR one-form in the non-Lorentzian supergravity that we will derive later. 
As expected, the truncation at $O(\tilde{\omega})$ should give a non-Lorentzian supergravity that matches the null reduction of eleven-dimensional supergravity. We have derived this null-reduced action~\eqref{eq:expsiia2} by taking the infinite $\omega$ limit of IIA supergravity. It can be seen that the reparametrization of $C^{(1)}$ is consistent with this expectation, which corroborates our previous guess in Section~\ref{sec:nlr}. 

Next, we generalize the $\tilde{\omega}$-expansion to include the Kalb-Ramond and RR potentials. In order to recover the same action from the large $\tilde{\omega}$ expansion, we find that a further rescaling of the RR potentials in $\tilde{\omega}$ is necessary, with 
$c^{(q)} \rightarrow \tilde{\omega}^{-3/2} \, c^{(q)}$. We are thus led to the following prescription:
\begin{subequations} \label{eq:m0tcbn}
\begin{align}
    G^{}_{\mu\nu} &= - \tilde{\omega} \, \omega \, \tau^{}_\mu \, \tau^{}_\nu + \frac{\tilde{\omega}}{\omega} \, e^{}_\mu{}^i \, e^{}_\nu{}^i,
        &%
    C^{(1)} &= \frac{\omega^2}{\tilde{\omega}} \, e^{-\varphi} \, \tau^{}_\mu \, \dd x^\mu + \frac{c^{(1)}}{\tilde{\omega}}, \\[4pt]
    B^{(2)} &= \tilde{\omega} \, b^{(2)},
        \qquad%
    \Phi = \varphi + \tfrac{3}{2} \ln \bigl( \tilde{\omega} / \omega \bigr)\,,
        &%
    C^{(3)} &= c^{(3)}.
\end{align}
\end{subequations}
In general, we have $C^{(q)} \sim \tilde{\omega}^{\frac{q-3}{2}} \, c^{(q)}$. It is understood that the Hodge duals between RR potentials are taken into account.
Couple $N$ non-interacting D-particles to the IIA supergravity and then expand with respect to both large $\omega$ and large $\tilde{\omega}$\,, we find
\be
    S = \tilde{\omega} \, S^{}_\text{\scalebox{0.8}{NL}} + \frac{N}{\tilde{\omega}} \, S^{\text{\scalebox{0.8}{M0T}}}_{\text{\scalebox{0.8}{D0}}} + O\bigl( \omega^{-2}, \, \tilde{\omega}^{-3} \bigr)\,, 
\ee
with $S_\text{\scalebox{0.8}{NL}}$ the non-Lorentzian supergravity action~\eqref{eq:expsiia2} and $S^{\text{\scalebox{0.8}{M0T}}}_{\text{\scalebox{0.8}{D0}}}$ the non-relativistic D-particle action~\eqref{eq:d0a}. At the leading $O(\tilde{\omega})$ order, the D-particles are decoupled as long as $N \ll \tilde{\omega}^2$\,. 

Finally, due to the dilatation symmetry~\eqref{eq:ds} of the non-Lorentzian supergravity~\eqref{eq:expsiia2}, the leading order term in the large $\omega$ and $\tilde{\omega}$ expansion remains unchanged if we fix the dilatation such that $\omega = \tilde{\omega}$\,. Now, the prescription~\eqref{eq:m0tcbn} reduces to
\begin{subequations} \label{eq:m0tcbnso}
\begin{align}
    G^{}_{\mu\nu} &= - \tilde{\omega}^2 \, \tau^{}_\mu \, \tau^{}_\nu + e^{}_\mu{}^i \, e^{}_\nu{}^i,
        &%
    C^{(1)} &= \frac{\tilde{\omega}}{e^{\varphi}} \, \tau^{}_\mu \, \dd x^\mu + \frac{c^{(1)}}{\tilde{\omega}}, \\[4pt]
    B^{(2)} &= \tilde{\omega} \, b^{(2)},
        \qquad\,\,%
    \Phi = \varphi\,,
        &%
    C^{(3)} &= c^{(3)}.
\end{align}
\end{subequations}
This ansatz obtained using M0T extends the original prescription~\eqref{eq:ftoa} that we have derived from the gravity dual to include the gauge potentials.\,\footnote{The one-form gauge potential on the D-brane should acquire the same rescaling as the Kalb-Ramond field due to gauge symmetry.}

We now apply the full prescription~\eqref{eq:m0tcbnso} to revisit the couplings between the non-Lorentzian supergravity and string/brane objects in the gravity dual, which was already considered in Section~\ref{sec:mln} but in absence of gauge potentials. We complete the string and brane actions in Eqs.~\eqref{eq:efodp} and \eqref{eq:nsf} to be
\begin{subequations} \label{eq:nsfs}
\begin{align}
    S^{}_{\text{\scalebox{0.8}{F1}}} &= \tilde{\omega} \, \frac{N}{\tilde{\ell}^{\,2}_\text{s}} \lr - \int \dd^2 \sigma \, \sqrt{- \det
		\CM_2} - \int b^{(2)} \rr + O\bigl( \tilde{\omega}^{-1} \bigr), \\[4pt]
    S^{}_{\text{\scalebox{0.8}{D$p$}}} &= \tilde{\omega} \, \frac{N}{\tilde{\ell}^{\,p+1}_\text{s}} \ls - \int \dd^{p+1} \sigma \, e^{-\varphi} \sqrt{- \det \CM_{p+1}} + \tilde{\omega}^{\frac{p-4}{2}} \! \int c^{(p+1)} \rs + O\bigl(\tilde{\omega}^{-1}\bigr)\,, \\[4pt]
    S^{}_\text{\scalebox{0.8}{NS5}} &= \tilde{\omega} \, \frac{N}{\tilde{\ell}^{\,6}_\text{s}} \ls - \int \dd^6 \sigma \, e^{-2\varphi} \, \sqrt{- \det \CM_{6}} + \tilde{\omega} \int b^{(6)} \rs + O\bigl(\tilde{\omega}^{-1}\bigr)\,, 
\end{align}
\end{subequations}
where $\CM_{n}$ is defined in Eq.~\eqref{eq:sdps}. Here, we used $\dd B^{(6)} = \star \bigl( e^{-2\Phi} \, \dd B^{(2)} \bigr)$ to obtain the $\tilde{\omega}$ dependence in front of $b^{(6)}$. 
It is manifest that only the static F1-string and D4-brane couple to the non-Lorentzian supergravity at the $O(\tilde{\omega})$ order, which matches the $\tilde{\omega}$ scaling in front of $b^{(2)}$ and $c^{(5)}$ in Eq.~\eqref{eq:nsfs}. 
Consequently, the same non-Lorentzian F1-string and D4-brane geometries that we have seen in Section~\ref{sec:bdcnlg} also exist at the $O(\tilde{\omega})$ order of the bulk gravity dual. Consider the flat spacetime version of the reparametrization~\eqref{eq:m0tcbnso},
\begin{align} \label{eq:m0tcbnf}
    t \rightarrow \tilde{\omega} \, t,
        \qquad%
    x^i \rightarrow x^i,
        \qquad%
    g^{}_\text{s} \rightarrow g^{}_\text{s}\,,
\end{align}
which we can apply asymptotically to the Lorentzian bulk geometries in order to derive curved non-Lorentzian geometries that match the M0T solutions in Section~\ref{sec:bdcnlg}: 
\begin{itemize}

\item

The F1-D0 geometry in the Lorentzian IIA theory is 
\begin{subequations} \label{eq:fodz}
\begin{align}
    \dd s^2_{10} &= \frac{1}{H_\text{\scalebox{0.8}{F1}}} \lr - \frac{\dd t^2}{\sqrt{H_\text{0}}} + \sqrt{H_\text{0}} \, \dd x_1^2 \rr + \sqrt{H_\text{0}} \, \bigl( \dd r^2 + r^2 \, \dd\Omega^{2}_7 \bigr)\,, \\[4pt]
    e^\Phi &= H^{3/4}_\text{0} \, \frac{g_{s}}{\sqrt{H_\text{\scalebox{0.8}{F1}}}}\,,
        \qquad%
    C^{(1)} = \frac{\dd t}{g^{}_\text{s} \, H_\text{0}}\,,
        \qquad%
    B^{(2)} = - \frac{\dd t \wedge \dd x^1}{H_\text{\scalebox{0.8}{F1}}}\,,
\end{align}
\end{subequations}
with $r^2 = x_2^2 + \cdots + x_9^2$\,. Here, $H_\text{0} = 1 + \ell_\text{0}^{\,6} / r^6$ and $H_\text{\scalebox{0.8}{F1}} = 1 + \ell_\text{\scalebox{0.8}{F1}}^{\,6} / r^6$ are the harmonic functions for the D0-branes and F-strings, respectively, with $\ell_\text{0}^{\,6} \sim N^{}_\text{0} \, g^{}_\text{s} \, \ell_\text{s}^7 / L_\text{\scalebox{0.8}{F1}}$\,, $L_\text{\scalebox{0.8}{F1}}$ the size of the compact $x^1$ direction, and $\ell_\text{\scalebox{0.8}{F1}}^{\,6} \sim N^{}_\text{\scalebox{0.8}{F1}} \, g_\text{s}^2 \, \ell_\text{s}^{\,6}$. 
Plugging the ansatz~\eqref{eq:m0tcbnf} into Eq.~\eqref{eq:fodz} and then comparing with Eq.~\eqref{eq:m0tcbnso} gives
\begin{subequations}
\begin{align} 
    \tau^{}_\mu \, \dd x^\mu &= H^{- 1/4}_\text{0} \, \frac{\dd t}{\sqrt{H_\text{\scalebox{0.8}{F1}}}}\,,
        &%
    e^{}_{\mu\nu} \, \dd x^\mu \, \dd x^\nu &= \sqrt{H^{}_\text{0}} \, \lr \frac{\dd x_1^2}{H_\text{\scalebox{0.8}{F1}}} + \dd r^2 + r^2 \, \dd \Omega_7^2 \rr, \\[4pt]
    e^\varphi &= H^{3/4}_\text{0} \, \frac{g^{}_\text{s}}{\sqrt{H_\text{\scalebox{0.8}{F1}}}}\,,
        &%
    b^{(2)} &= - \frac{\dd t \wedge \dd x^1}{H_\text{\scalebox{0.8}{F1}}}\,,
\end{align}
\end{subequations}
This is a solution to the truncated non-Lorentzian supergravity at the $O(\tilde{\omega})$ order. As expected, this is equivalent to the M0T geometry~\eqref{eq:tef} up to a dilatation transformation~\eqref{eq:ds} with $\Omega = \sqrt{H^{}_\text{0}}$\,. 

\item

The D0-D4 geometry in the Lorentzian  IIA theory is
\begin{align} \label{eq:fodf}
    \dd s^2_{10} &= - \frac{\dd t^2}{\sqrt{H^{}_\text{0} \, H^{}_\text{4}}} + \sqrt{\frac{H^{}_\text{0}}{H^{}_\text{4}}} \, \bigl( \dd x_1^2 + \cdots + \dd x_4^2 \bigr) + \sqrt{H^{}_\text{0} \, H^{}_\text{4}} \, \bigl( \dd r^2 + r^2 \, \dd \Omega_4^2 \bigr)\,, \notag \\[4pt]
    e^\Phi &= H^{3/4}_\text{0} \, \frac{g_\text{s}}{H^{1/4}_\text{4}}\,,
        \qquad%
    C^{(1)} = \frac{\dd t}{g^{}_\text{s} \, H^{}_\text{0}}\,,
        \qquad%
    C^{(5)} = \frac{\dd t \wedge \cdots \wedge \dd x^4}{g_\text{s} \, H^{}_\text{4}}\,.
\end{align}
Here, $H_\text{0} = 1 + \ell_\text{0}^3 / r^3$ and $H_\text{4} = 1 + \ell_\text{4}^3 / r^3$ are the harmonic functions for the D0- and D4-branes, respectively, with $\ell_\text{0}^3 \sim N_\text{0} \, g^{}_\text{s} \, \ell_\text{s}^7 / V_4$\,, $V_4$ the volume of the volume of the
four-dimensional internal compact manifold along $x_1, \cdots, x_4$ wrapped by the D4-branes, and $\ell_\text{4}^3 \sim N_\text{4} \, g_\text{s} \, \ell_\text{s}^3$.  
Plugging the ansatz~\eqref{eq:m0tcbnf} and then comparing with Eq.~\eqref{eq:m0tcbnso} gives 
\begin{subequations}
\begin{align} 
    & \tau^{}_\mu \, \dd x^\mu = H^{- 1/4}_\text{0} \, \frac{\dd t}{H_\text{4}^{1/4}}\,,
        \quad%
    e^\varphi = H^{3/4}_\text{0} \, \frac{g^{}_\text{s}}{H_\text{4}^{1/4}}\,,
        \quad%
    c^{(5)} = \frac{1}{g^{}_\text{s} \, H^{}_\text{4}} \, \dd t \wedge \cdots \wedge \dd x^4, \\[4pt]
    & e^{}_{\mu\nu} \, \dd x^\mu \, \dd x^\nu = \sqrt{H^{}_\text{0}} \ls \frac{\dd x_1^2 + \cdots + \dd x_4^2}{\sqrt{H^{}_\text{4}}} + \sqrt{H^{}_\text{4}} \, \bigl( \dd r^2 + r^2 \, \dd \Omega_4^2 \bigr) \rs, 
\end{align}
\end{subequations}
This is equivalent to Eq.~\eqref{eq:m0td4} up to a dilatation transformation~\eqref{eq:ds} with $\Omega = \sqrt{H^{}_\text{0}}$\,. 

\end{itemize}

\subsection{Generalizations to D\texorpdfstring{$p$}{p}-Brane Holography}

We now generalize the D-particle $\tilde{\omega}$-expansion to other D$p$-branes. In D$p$-brane holography~\cite{Itzhaki:1998dd}, matrix (gauge) theory at large $N$ can acquire a weak gravity dual on the near-horizon D$p$-brane geometry,
\begin{subequations} \label{eq:nhgp}
\begin{align} 
	\dd s^2_{10} &= \frac{-\dd t^2 + \cdots + \dd x_p^2}{\sqrt{h}} + \sqrt{h} \, \bigl( \dd r^2 + r^2 \, \dd \Omega^2_{8-p} \bigr), 
        		\qquad%
    	e^\Phi = g^{}_\text{s} \, h^{\frac{3-p}{4}}, \\[4pt]
	C^{(p+1)} &= \frac{\dd t \! \wedge \cdots \wedge \! \dd x^p}{g^{}_\text{s} \, h},
		\qquad%
    	h = \CC_{p} \, N \, g^{}_\text{s} \lr \frac{\ell^{}_\text{s}}{r} \rr^{\!\!7-p}\!,
		\qquad
	\CC_p = \bigl( 2 \sqrt{\pi} \bigr)^{5-p} \, \Gamma\bigl( \tfrac{7-p}{2} \bigr)\,,
\end{align}
\end{subequations}
which contains a (conformally) AdS sector. Here, $r^2 = x_{p+1}^2 + \cdots + x_9^2$\,.  
The weakly coupled gravity on such a background dominates if 
\be \label{eq:range}
    1 \ll g_\text{eff}^2 \ll N^{\frac{4}{7-p}},
        \qquad%
    g^2_\text{eff} \sim \frac{N g^2_\text{YM}}{E^{3-p}} \sim N g^{}_\text{s} \! \lr \frac{\ell^{}_\text{s}}{r} \rr^{\!\!3-p}\!,
\ee
with $g^2_\text{YM} \sim g^{}_\text{s} / \ell^{3-p}_\text{s}$ the YM coupling of the dual gauge theory and $ E \sim r/l_s^2$\,. Keeping $g^{}_\text{eff}$ unchanged such that the condition~\eqref{eq:range} is valid, we are led to the following generalization of Eq.~\eqref{eq:gslsrs0}:
\be \label{eq:todbr}
    g^{}_\text{s} = \tilde{\omega}^{\frac{3-p}{2}} \, \tilde{g}^{}_\text{s}\,,
        \qquad%
    \ell^{}_\text{s} = \tilde{\omega}^{-\frac{1}{2}} \, \tilde{\ell}^{}_\text{s}\,.
\ee
Under these rescalings, the near-horizon geometry~\eqref{eq:nhgp} becomes
\begin{subequations} \label{eq:nhgprs}
\begin{align} 
	\dd s^2_{10} &= \frac{\tilde{\omega}}{\tilde{h}^{\frac{1}{2}}} \bigl( -\dd t^2 + \cdots + \dd x_p^2 \bigr) + \frac{\tilde{h}^{\frac{1}{2}}}{\tilde{\omega}} \, \bigl( \dd r^2 + r^2 \, \dd \Omega^2_{8-p} \bigr)\,, 
        		&%
    e^\Phi &= \tilde{g}^{}_\text{s} \, \tilde{h}^{\frac{3-p}{4}}, \\[4pt]
	C^{(p+1)} &= \frac{\tilde{\omega}^{\frac{p+1}{2}}}{\tilde{g}^{}_\text{s} \, \tilde{h}} \, \dd t \! \wedge \cdots \wedge \! \dd x^p\,,
		&%
    \tilde{h} &= \CC_p \, N \, \tilde{g}^{}_\text{s} \lr \frac{\tilde{\ell}^{}_\text{s}}{r} \rr^{\!\!7-p}\!\!,
\end{align}
\end{subequations}
with $\CC_p$ an unimportant constant factor. Note that $e^\Phi$ is independent of $\tilde{\omega}$. 
Furthermore, the rescaling of the string length can be absorbed into extra rescalings of the background fields by using Eq.~\eqref{eq:rsgt}, which leads us to the following covariant prescription generalizing Eq.~\eqref{eq:ftoa}:
\begin{subequations} \label{eq:nhgp02}
\begin{align} 
	\dd s^2_{10} &= \tilde{\omega}^2 \, \tau^{}_{\mu\nu} + e^{}_{\mu\nu}\,, 
        		\qquad%
    	e^\Phi = e^{\varphi}, 
		\qquad%
	C^{(p+1)} = \tilde{\omega}^{\,p} \! \lr \frac{\tilde{\omega}}{e^\varphi} \, \tau^0 \wedge \cdots \wedge \tau^p + \frac{c^{(p+1)}}{\tilde{\omega}} \rr\!,
\end{align}
\end{subequations}
with $\tau^{}_{\mu\nu} = \tau^{}_\mu{}^a \, \tau^{}_\nu{}^b \, \eta^{}_{ab}$\,, $a = 0\,,\,\cdots,\,p$ and $e^{}_{\mu\nu} = e^{}_\mu{}^i \, e^{}_\nu{}^i$, $i = p+1\,,\,\cdots, 9$\,. The vacuum expectation values are 
\be \label{eq:vevgp}
    \langle \tau_\mu{}^a \rangle = \delta_\mu^a \, \tilde{h}^{-1/4}, 
        \qquad%
    \langle e_\mu{}^i \rangle = \delta_\mu^i \, \tilde{h}^{1/4}, 
        \qquad%
    \langle e^\varphi \rangle = \tilde{g}_\text{s} \, \tilde{h}^{(3-p)/4}. 
\ee    
At large $\tilde{\omega}$, in general we are led to an approximately non-Lorentzian regime described by the vielbein fields $\tau_\mu{}^a$ and $e_\mu{}^i$, encoding the longitudinal and transverse sectors, respectively. Unlike the particle case, the longitudinal sector is now ($p$+1)-dimensional instead of the one-dimensional time direction. In this sense, the vielbein fields describe a $p$-brane version of the Newton-Cartan geometry. Just like the D-particle case, a moderately large $N$ with $1 \ll N \ll \tilde{\omega}^2$ is required for this non-Lorentzian regime to be valid. 

On the matrix gauge theory side, the near-horizon limit giving rise to the bulk geometry~\eqref{eq:nhgp} corresponds to the BPS decoupling limit. Such a limiting prescription can be obtained by T-dualizing M0T prescription~\eqref{eq:m0tcb} on a spatial $n$-torus. This leads to the following limiting prescription for the Matrix $p$-brane Theory (M$p$T) as a corner of type II superstring theory: 
\begin{align} \label{eq:mptcb}
    G^{}_{\!\mu\nu} &= \omega \, \tau^{}_{\mu\nu} + \frac{e^{}_{\mu\nu}}{\omega}\,,
        &%
    C^{(p+1)} &= \frac{\omega^2}{e^{\varphi}} \, \tau^0 \! \wedge \cdots \wedge \! \tau^p + c^{(p+1)}, 
        &%
    \Phi = \varphi + \tfrac{p-3}{2} \, \ln \omega\,.
\end{align}
In the $\omega \rightarrow \infty$ limit, the fundamental degrees of freedom in M$p$T (at least for $p \leq 5$) are captured by the ($p$+1)-dimensional matrix (gauge) theory. The $\omega \rightarrow \infty$ limit of IIA/B supergravity typically leads to a non-Lorentzian supergravity,\,\footnote{Such a non-Lorentzian supergravity is T-dual to the M0T supergravity. Some related ingredients can be found in~\cite{Blair:2021waq, Blair:2023noj, Lambert:2024ncn, Bergshoeff:2024ipq}. We leave the detailed study of general M$p$T supergravity theories for future work.} which admits a Galilei-like boost symmetry that map the transverse to longitudinal directions. 
The emergent dilatation symmetry in the $\omega \rightarrow \infty$ limit acts as
\be \label{eq:dsdp}
	\tau^{}_\mu{}^a \rightarrow \Omega^{\frac{1}{2}} \, \tau^{}_\mu{}^a,
		\qquad%
	e^{}_\mu{}^i \rightarrow \Omega^{- \frac{1}{2}} \, e^{}_\mu{}^i,
		\qquad%
	e^\Phi \rightarrow \Omega^{\frac{p-3}{2}} \, e^\varphi.
\ee

Similar to the D-particle case, backreacting the fundamental D$p$-branes deforms the associated non-Lorentzian supergravity to the full IIA/B theory on the near-horizon geometry~\eqref{eq:nhgp}. Therefore, when the D$p$-brane backreaction is taken into account, we should consider an $\omega$-expansion instead of the strict $\omega \rightarrow \infty$ limit. Furthermore, introducing the $\tilde{\omega}$-expansion decouples the fundamental D$p$-branes at the leading order. At moderately large $N$ with $1 \ll N \ll \tilde{\omega}^2$\,, we are left with the leading-order contribution as classical non-Lorentzian supergravity. Just like in the D-particle case, this identifies the non-Lorentzian supergravity that corresponds to the truncation at the leading $O(\tilde{\omega})$ order on the bulk side. 

Comparing with the M$p$T supergravity, we infer the generalization of the $\tilde{\omega}$-expansion prescription~\eqref{eq:m0tcbn} to the D$p$-brane case, with
\begin{subequations} \label{eq:mptcbn}
\begin{align}
    G^{}_{\!\mu\nu} &= \tilde{\omega} \, \omega \, \tau^{}_{\mu\nu} + \frac{\tilde{\omega}}{\omega} \, e^{}_{\mu\nu}\,,
        &%
    C^{(p+1)} &= \tilde{\omega}^{\,p-1} \! \lr \omega^2 \, e^{-\varphi} \, \tau^0 \! \wedge \! \cdots \! \wedge \! \tau^p + c^{(p+1)} \rr, \\[4pt]
    \Phi &= \varphi + \tfrac{3 - p}{2} \ln \bigl( \tilde{\omega} / \omega \bigr)\,,
        &%
    B^{(2)} &= \tilde{\omega} \, b^{(2)},
            \qquad%
    C^{(q)} = \tilde{\omega}^{\frac{p + q - 3}{2}} \, c^{(q)}, 
    	\,\,\,\,
    q \neq p+1\,.
\end{align}
\end{subequations}
It is understood that Hodge dual conditions relating the RR potentials $C_{q}$ and $C_{8-q}$ should be taken into account. 
Using the dilatation transformation~\eqref{eq:dsdp} at the leading order, we set $\omega = \tilde{\omega}$\,, which gives rise to the following expansion prescription that does not alter the leading-order term: 
\begin{subequations} \label{eq:mptcbnf}
\begin{align} 
    	G^{}_{\!\mu\nu} &= \tilde{\omega}^2 \, \tau^{}_{\mu\nu} + e^{}_{\mu\nu}\,, 
    		&%
	C^{(p+1)} &= \tilde{\omega}^{\,p-1} \! \lr \tilde{\omega}^2 \, e^{-\varphi} \, \tau^0 \! \wedge \! \cdots \! \wedge \! \tau^p + c^{(p+1)} \rr, \\[4pt]
	B^{(2)} &= \tilde{\omega} \, b^{(2)},
        		\quad%
    	\Phi = \varphi\,,
        		&%
    	C^{(q)} &= \tilde{\omega}^{\frac{p + q - 3}{2}} \, c^{(q)},
		\quad
	q \neq p+1\,.
        \label{eq:mptcbnfs}
\end{align}
\end{subequations}
In the associated truncated non-Lorentzian supergravity, the dilation symmetry implies that the vacuum expectation values in Eq.~\eqref{eq:vevgp} are equivalent to flat background. 

Next, we consider the backreaction of static extended objects in the truncated non-Lorentzian supergravity at the leading $\tilde{\omega}$ order. The effective gravitational coupling is 
\be
    \kappa^2_\text{eff} \,\, \propto \,\, g^2_\text{s} \, \ell^{\,8}_\text{s} = \tilde{\omega}^{-p-1} \, \tilde{g}^{\,2}_\text{s} \, \tilde{\ell}^{\,8}_\text{s}\,.
\ee
Hence, the expansion of the type II supergravity action with respect to large $\tilde{\omega}$ gives
\be \label{eq:gsg}
    S^{}_\text{sugra} = \tilde{\omega}^{\,p+1} \, S^{\text{\scalebox{0.8}{D$p$}}}_\text{\scalebox{0.8}{NL}} + O\bigl( \tilde{\omega}^{\,p-1} \bigr)\,.
\ee
Here, $S^{\text{\scalebox{0.8}{D$p$}}}_\text{\scalebox{0.8}{NL}}$ coincides with the M$p$T limit of type II supergravity action. A static D$q$-brane source that backreacts to form a solution to this supergravity must satisfy
\be
    \int C^{(q+1)} = \tilde{\omega}^{\,p+1} \int c^{(q+1)} 
        \quad\implies\quad%
    q = p+4\,,
\ee
where we have used the second equation in Eq.~\eqref{eq:mptcbnfs}. This implies that there is a D($p$+4)-brane solution in the truncated supergravity. Such a solution also naturally arises in the context of M$p$T supergravity, as D$p$-D($p$+4) is a $\frac{1}{4}$-BPS state. Using the non-Lorentzian geometric data in Eq.~\eqref{eq:mptcbnf}, this non-Lorentzian D($p$+4)-brane geometry is ($p \leq 2$)
\begin{align} \label{eq:mptd4}
    \tau^{}_{\mu\nu} \, \dd x^\mu \, \dd x^\nu &= \frac{-\dd t^2 + \cdots + \dd x^2_p}{\sqrt{H}}\,, 
        \qquad\quad\,%
    e^\varphi = \frac{g^{}_\text{s}}{H^{(p+1)/4}}\,, 
        &%
    c^{(p+5)} &= \frac{\dd t \wedge \cdots \wedge \dd x^{p+5}}{g^{}_\text{s} \, H}\,, \notag \\[4pt]
    e^{}_{\mu\nu} \, \dd x^\mu \, \dd x^\nu &= \frac{\dd x_{p+1}^2 + \cdots + \dd x_{p+4}^2}{\sqrt{H}} + \sqrt{H} \, \bigl( \dd r^2 + r^2 \, \dd \Omega_{4-p}^2 \bigr)\,, 
        &%
    H &= 1 + \lr \frac{\ell}{r} \rr^{\!\!3-p}\!.
\end{align}
When $p=3$, \emph{i.e.}~the classic case of the AdS${}_5/$CFT${}_4$ correspondence, we encounter the non-Lorentzian D7-brane geometry as a BPS solution, where the harmonic function develops a logarithmic dependence. Such an exotic brane geometry typically needs to be completed at large $r$ and further analysis is required, which would be interesting to study in the future. 
The brane solution becomes more exotic for $p>3$\,, which we do not pursue here.
On the other hand, a static F-string source term always scales as $O(\tilde{\omega})$\,, and it is only sourcing the leading-order non-Lorentzian supergravity in Eq.~\eqref{eq:gsg} when $p=0$\,, \emph{i.e.}~the D-particle case. For NS5-brane, the static source term scales as $O(\tilde{\omega}^{p+2})$ and does not match the leading $O(\tilde{\omega}^{p+1})$ order in Eq.~\eqref{eq:gsg}.

\subsection{String Soliton Holography}

Another interesting corner to zoom in on its non-Lorentzian edge is associated with matrix string theory and its dual gravity on the string soliton background. The near-horizon string soliton solution is
\be \label{eq:tnhfst}
    \dd s^2_{10} = \frac{- \dd t^2 + \dd x_1^2}{h} + \dd r^2 + r^2 \, \dd\Omega_7^2\,,
        \quad%
    B^{(2)} = - \frac{\dd t \wedge \dd x^1}{h}\,,
        \quad%
    e^\Phi = \frac{g^{}_\text{s}}{\sqrt{h}}\,,
        \quad%
    h = \frac{\ell^6}{r^6}\,,
\ee
where $\ell^6 = 32 \, \pi^2 \, w \, g^2_\text{s} \, \ell^{6}_\text{s}$\,, with $w$ the string winding number. The weakly coupled gravity dominates if $1 \ll r / \ell_\text{s} \ll w^{1/6}$~\cite{Itzhaki:1998dd, Harmark:2025ikv}.
This range opens up at large winding number. 
This condition holds as long as $\ell_\text{s} = \tilde{\ell}_\text{s}$ remains untouched, under a rescaling analogous to Eq.~\eqref{eq:todbr} in the D-brane case. This leaves the rescaling of $\tilde{g}^{}_\text{s}$ free, which we take to be $g^{}_\text{s} = \tilde{g}^{}_\text{s} / \tilde{\omega}$\,. Under this rescaling, the near-horizon geometry~\eqref{eq:tnhfst} becomes
\begin{subequations} \label{eq:nhgprs2}
\begin{align} 
	\dd s^2_{10} &= \frac{\tilde{\omega}^2}{\tilde{h}} \Bigl( -\dd t^2 + \dd x_1^2 \Bigr) + \dd r^2 + r^2 \, \dd \Omega^2_{7}\,, 
        		&\hspace{-3cm}%
    e^\Phi = \frac{\tilde{g}^{}_\text{s}}{\tilde{h}^{\frac{1}{2}}}\,,& \\[4pt]
	B^{(2)} &= - \tilde{\omega}^2 \, \frac{\dd t \wedge \dd x^1}{\tilde{h}}\,,
		&\hspace{-3cm}%
    \tilde{h} = 32 \, \pi^2 \, w \, \tilde{g}^{2}_\text{s} \lr \frac{\tilde{\ell}^{}_\text{s}}{r} \rr^{\!\!\!6}\!,&
\end{align}
\end{subequations}
which gives rise to the covariant prescription,
\begin{subequations} \label{eq:nhgp0}
\begin{align} 
	\dd s^2_{10} &= \tilde{\omega}^2 \, \tau^{}_{\mu\nu} + e^{}_{\mu\nu}\,, 
        		\qquad%
    	e^\Phi = e^{\varphi}, 
		\qquad%
	B^{(2)} = \tilde{\omega}^2 \, \tau^0 \wedge \tau^1 + b^{(2)},
\end{align}
\end{subequations}
with $\tau^{}_{\mu\nu} = \tau^{}_\mu{}^a \, \tau^{}_\nu{}^b \, \eta^{}_{ab}$\,, $a = 0\,,\,1$ and $e^{}_{\mu\nu} = e^{}_\mu{}^i \, e^{}_\nu{}^i$, $i = 2\,,\,\cdots, 9$\,. The background values are 
\be \label{eq:vevgps}
    \langle \tau_\mu{}^a \rangle = \delta_\mu^a \, \tilde{h}^{-1/2}, 
        \qquad
    \langle e_\mu{}^i \rangle = \delta_\mu^i\,, 
        \qquad%
    \langle e^\varphi \rangle = \tilde{g}^{}_\text{s} / \tilde{h}^{1/2}. 
\ee  
The gravitational constant now scales as
$\kappa^2_\text{eff} \,\, \propto \,\, g^2_\text{s} \, \ell^{\,8}_\text{s} = \tilde{\omega}^{-2} \, \tilde{g}^{\,2}_\text{s} \, \tilde{\ell}^{\,8}_\text{s}$\,. It is required that the string winding $w$ is moderately large such that $1 \ll w \ll \tilde{\omega}^2$.
Focusing on the type IIB case for matrix string theory, the large $\tilde{\omega}$ expansion of the supergravity takes the form,
\be \label{eq:sefs}
    S^{}_\text{sugra} = \tilde{\omega}^{2} S^{\text{\scalebox{0.8}{F1}}}_\text{\scalebox{0.8}{NL}} + O\bigl( \tilde{\omega}^{\,0} \bigr)\,,
\ee
where the truncated non-Lorentzian supergravity described by $S^{\text{\scalebox{0.8}{F1}}}_\text{\scalebox{0.8}{NL}}$ has been studied in~\cite{Bergshoeff:2023ogz}. Meanwhile, the leading order term of the F-string action vanishes, which implies that there is no backreaction from the F-string in the truncated non-Lorentzian supergravity. 
Matching onto the string Newton-Cartan supergravity~\cite{Bergshoeff:2023ogz} that emerges in the target space from non-relativistic string theory, we infer that the RR potentials are reparametrized as
\be
    C^{(q)} = \tilde{\omega} \Bigl( \tilde{\omega}^2 \, e^{-\varphi} \, \tau^0 \wedge \tau^1 \wedge c^{(q-2)} + c^{(q)} \Bigr)\,.
\ee
The static source term is again subleading compare to the leading $O(\tilde{\omega}^2)$ order in 
Eq.~\eqref{eq:sefs}. 

An interesting source that couples to this truncated stringy non-Lorentzian supergravity is the NS5-brane, Using the standard supergravity equation of motion $\dd B^{(6)} = \star \bigl( e^{-2\Phi} \, \dd B^{(2)} \bigr)$\,, we infer that the static NS5-brane source takes the form $\tilde{\omega}^2 \int b^{(6)}$, which matches the non-Lorentzian supergravity term in Eq.~\eqref{eq:sefs}. The associated NS5-brane solution in the truncated non-Lorentzian supergravity is
\begin{subequations}
\begin{align}
    \tau^{}_{\mu\nu} \, \dd x^\mu \, \dd x^\nu &= - \dd t^2 + \dd x_1^2,
        &%
    e^{\Phi} &= g^{}_\text{s} \, \sqrt{H}\,, \\[4pt]
    e^{}_{\mu\nu} \, \dd x^\mu \, \dd x^\nu &= \dd x_2^2 + \cdots + \dd x_5^2 + H \, \bigl( \dd r^2 + r^2 \, \dd \Omega^2_3 \bigr)\,,
        &%
    \bigl(\dd b^{(2)}\bigr)_{ijk} &= \epsilon_{ijk\ell} \, \p^{}_\ell H\,,
\end{align}
\end{subequations}
where $H = 1 + N \ell^2_\text{s} / r^2$\,. This coincides with the non-Lorentzian NS-brane solution in~\cite{Harmark:2025ikv}. 

\section{Outlook}

In this paper we studied different non-Lorentzian regimes related to matrix theory. We clarified the role of non-Lorentzian supergravity in the context of D-brane and string soliton holography, and studied the sourcing in such a novel gravitational system by extended objects in string theory. This study can be viewed as a further step towards a comprehensive understanding of the BPS decoupling limits of string theory and M-theory, with rich but more tractable physics closely related to the AdS/CFT correspondence and matrix theory. 

In a next step, it would be important to analyze the general non-Lorentzian supergravity theories related to matrix (gauge) theories~\cite{Blair:2021waq, Blair:2023noj, Lambert:2024ncn, Bergshoeff:2024ipq}, including the fermionic sector. This is key to determining the status of potential geometric constraints (see Appendix~\ref{app:pnrst}). A series of studies of supergravities in the context of non-relativistic string theory and M-theory have been performed in~\cite{Bergshoeff:2018vfn, Gallegos:2020egk, Bergshoeff:2021bmc, Bergshoeff:2021tfn, Blair:2021waq, Bergshoeff:2023ogz, Bergshoeff:2024ipq, Bergshoeff:2025grj}, which provide useful ingredients for their generalizations to other non-Lorentzian supergravities associated with the BPS decoupling limits zooming in on D-branes instead of the F-string. 
A different avenue to pursue is along the lines of~\cite{VandenBleeken:2017rij, Hansen:2019pkl, Hansen:2020pqs, Hansen:2020wqw, Hartong:2021ekg, Hartong:2022dsx, Hartong:2024ydv} to analyze the structure of the higher-order terms in a covariant $\tilde{\omega}$-expansion of supergravity. 
It would also be interesting to look for solutions with blackening factors, along the lines of \emph{e.g.}~\cite{Blair:2025ewa} and beyond. This would allow us to extend our discussions to the thermal case. Finally, how precisely such non-Lorentzian supergravity theories emerge from ambitwistor-like string worldsheet theory as explained in Section~\ref{sec:nlgsw} deserves further investigations.  

Although we focused on Galilei-like supergravities that exhibit BPS behaviors, there are also other interesting corners in which the target space geometry is Carroll-like~\cite{Blair:2025nno}. For examples, in Section~\ref{sec:nlgsw} we have mentioned that a spatial T-duality transformation of the IIB${}^*$ supergravity associated with the IKKT matrix theory is Carroll-like~\cite{Blair:2023noj, Gomis:2023eav}. This class of Carroll-like geometries are also known to arise in the asymptotic regime of a bulk geometry containing a (conformally) de Sitter sector~\cite{Blair:2025nno, Argandona:2025jhg}. The discussion in Section~\ref{sec:nlgsw} then implies that there should be a worldsheet formalism for the dynamics of such Carroll-like supergravities. On the other hand, it is interesting to note that, in the D-brane and string soliton holography, the supergravity regime continues to hold not only at large $\tilde{\omega}$ but also small $\tilde{\omega}$. In this latter case, we are approaching a different Carrollian regime akin to the one discussed in~\cite{Blair:2025nno} associated with AdS space. This might provide us with a matrix theoretic perspective for understanding the dynamics in Carroll-like gravity. 

\acknowledgments

We would like to thank Eric Bergshoeff, Paolo Di Vecchia, Troels Harmark, Jelle Hartong, Xiaoan Lin, Niels A. Obers, L\'{a}rus Thorlacius for useful discussions. Z.Y. would like to thank the organizers and participants of the workshop \emph{Matrix Model for Superstring/M-Theory} at the Yukawa Institute in December 2025 for stimulating discussions.
The work of Z.Y. is supported by the Villum Young Investigator Programme under project No.~71589 and Olle Engkvists Stiftelse Project Grant 234-0342. The Center of Gravity is a Center of Excellence funded by the Danish National Research Foundation under grant No.~184.  U.Z. is supported by the Scientific and Technological Research Council of Türkiye (TÜBİTAK) under Grant Nos. 125F467 (ARDEB 1001) and 125F024 (ARDEB 3501)\,. 

\appendix

\section{Variation of Non-Lorentzian Supergravity} \label{app:eom}

We collect the equations of motion from varying the non-Lorentzian supergravity action~\eqref{eq:expsiia2} with respect to different fields below. It is understood that all derivatives are given by $\dd_\mu$ as defined in Eq.~\eqref{eq:dmo}, which makes the invariance under the dilatation transformation~\eqref{eq:ds} manifest. The notation $[\CO]$ denotes the variation of the action with respect to the field $\CO$, and the subscripts $0$ (and $i$) denotes the project by contracting with $\tau^\mu$ ($e^\mu{}_i$). We also define
$t^{}_{\mu\nu\rho} \equiv \nabla^{}_{\!\mu} t^{}_{\nu\rho}$\,, with $t^{}_{\mu\rho} = \dd_{[\mu} \tau_{\nu]}$ the torsion introduced in Eq.~\eqref{eq:ts}, and the extrinsic curvature 
$k_{\mu\nu} = \tfrac{1}{2} \, \tau^\rho \, ( \partial_\rho e_{\mu\nu} - \partial_\mu e_{\rho\nu} - \partial_\nu e_{\mu\rho} )$\,.   
The variations of the action~\eqref{eq:expsiia2} are
\begin{subequations}
\begin{align}
[\tau]_0 =\,&
e^{-2\varphi}
\bigl(
-r_{ii}
+ 2\,t_{0i}\,t_{0i}
- 2\,t_{i0i}
+ \tfrac14\,h_{0ij}\,h_{0ij}
\bigr)
+ 2\,e^{-\varphi}\,t_{ij}\,f_{ij}
+ \tfrac1{48}\,\widetilde f_{ijkl}\,\widetilde f_{ijkl}
\,, \\[6pt]
[\tau]_i =\,&
2\,e^{-2\varphi}
\bigl(
t_{jij}
- 2\,t_{0j}\,t_{ij}
+ \tfrac14\,h_{ijk}\,h_{0jk}
\bigr)
+ \tfrac16\,e^{-\varphi}\,\widetilde f_{ijkl}\,h_{jkl}
\,, \\[6pt]
[e]_{0i} =\,&
e^{-2\varphi}
\bigl(
r_{i0}
- k_{jj}\,t_{0i}
+ k_{ij}\,t_{0j}
- k_{0j}\,t_{ij}
+ t_{00i}
\bigr)
- e^{-\varphi}
\bigl(
2\,t_{0j}\,f_{ij}
+ t_{ij}\,f_{0j}
\bigr)
\notag \\[4pt]
&+
\tfrac12\,e^{-\varphi}
\bigl(
\nabla_{\!j} f_{ij}
+ \tfrac12\,h_{0jk}\,\widetilde f_{0ijk}
\bigr)
- \tfrac1{12}\,\widetilde f_{0jkl}\,\widetilde f_{ijkl}
\,, \\[6pt]
[e]_{ij} =\,&
e^{-2\varphi}
\bigl(
r_{ij}
+ k_{ik}\,t_{jk}
- k_{jk}\,t_{ik}
+ k_{kk}\,t_{ij}
- 2\,t_{ji0}
+ t_{0ij}
- 2\,t_{0i}\,t_{0j}
+ \tfrac12\,h_{0ik}\,h_{0jk}
\bigr)
\notag \\[4pt]
&+
\tfrac14\,e^{-\varphi}
\bigl(
h_{ikl}\,\widetilde f_{0jkl}
+ h_{jkl}\,\widetilde f_{0ikl}
+ 8\,t_{k(i}\,f_{j)k}
\bigr)
- \tfrac1{12}\,\widetilde f_{iklm}\,\widetilde f_{jklm}
\notag \\[4pt]
&+
\delta_{ij}\,e^{-2\varphi}
\bigl(
-\tfrac12\,r_{kk}
- 2\,t_{k0k}
+ 3\,t_{0k}\,t_{0k}
- \tfrac18\,h_{0kl}\,h_{0kl}
\bigr)
\notag \\[4pt]
&+
\delta_{ij}
\bigl(
\tfrac12\,e^{-\varphi}\,t_{kl}\,f_{kl}
- \tfrac1{12}\,e^{-\varphi}\,h_{klm}\,\widetilde f_{0klm}
+ \tfrac1{96}\,\widetilde f_{klmn}\,\widetilde f_{klmn}
\bigr)
\,, \\[6pt]
[\varphi] =\,&
-2\,r_{ii}
- 10\,t_{i0i}
+ 16\,t_{0i}\,t_{0i}
- \tfrac12\,h_{0ij}\,h_{0ij}
+ e^\varphi
\bigl(
f_{ij}\,t_{ij}
- \tfrac16\,\widetilde f_{0ijk}\,h_{ijk}
\bigr)
\,, \\[6pt]
[b]_{0i} =\,&
-\tau^\mu\,\nabla_{\!j}h_{\mu ij}
+ e^\varphi
\bigl(
t_{jk}\,\widetilde f_{0ijk}
+ \tfrac12\,h_{ijk}\,f_{jk}
\bigr)
-
\tfrac{e^{2\varphi}}{2(4!)^2}\,
\varepsilon^{i j_1\cdots j_8}\,
\widetilde f_{j_1j_2j_3j_4}\,
\widetilde f_{j_5j_6j_7j_8}
\,, \\[6pt]
[b]_{ij} =\,&
-\tau^\mu\,\nabla_{\!0}h_{\mu ij}
- k_{ik}\,h_{0jk}
+ k_{jk}\,h_{0ik}
- k_{kk}\,h_{0ij}
- e^\varphi
\bigl(
h_{ijk}\,f_{k0}
- 2\,t_{0k}\,\widetilde f_{0ijk}
\bigr)
\notag \\[4pt]
&-
\tfrac12\,e^{2\varphi}\,f_{kl}\,\widetilde f_{ijkl}
- e^\varphi\,\nabla_{\!k}\widetilde f_{0kij}
- \tfrac{e^{2\varphi}}{3!\,4!}\,
\varepsilon^{ij k_1\cdots k_7}\,
\widetilde f_{0k_1k_2k_3}\,
\widetilde f_{k_4k_5k_6k_7}
\,, \\[6pt]
[c]_0 =\,&
-2\,e^{-\varphi}
\bigl(
t_{ij}\,t_{ij}
- \tfrac1{12}\,h_{ijk}\,h_{ijk}
\bigr)
\,,
\label{eq:eomc0}
\\[6pt]
[c]_i =\,&
e^{-\varphi}
\bigl(
-2\,t_{jij}
- \tfrac12\,h_{0jk}\,h_{ijk}
+ 4\,t_{ij}\,t_{0j}
\bigr)
- \tfrac16\,h_{jkl}\,\widetilde f_{ijkl}
\,, \\[6pt]
[c]_{0ij} =\,&
e^{-\varphi}\,\nabla_{\!k}h_{ijk}
- t_{kl}\,\widetilde f_{ijkl}
- \tfrac1{3!\,4!}\,
\varepsilon^{ij k_1\cdots k_7}\,
h_{k_1k_2k_3}\,
\widetilde f_{k_4k_5k_6k_7}
\,,
\label{eq:eomc0ij}
\\[6pt]
[c]_{ijk} =\,&
-e^{-\varphi}
\bigl(
\nabla_{\!0}h_{ijk}
- k_{i\ell}\,h_{jk\ell}
+ k_{j\ell}\,h_{ik\ell}
- k_{k\ell}\,h_{ij\ell}
+ k_{\ell\ell}\,h_{ijk}
\bigr)
-
\nabla_{\!\ell}\widetilde f_{ijk\ell}
- 2\,t_{0\ell}\,\widetilde f_{\ell ijk}
\notag \\[4pt]
&+
\tfrac1{2\times4!}\,e^{-1}\,
\nabla_{\mu_0}
\bigl(
\varepsilon^{\mu_0\cdots\mu_9}\,
b_{\mu_1\mu_2}\,
f_{\mu_3\mu_4\mu_5\mu_6}
\bigr)\,
e_{\mu_7 i}\,e_{\mu_8 j}\,e_{\mu_9 k}
\,.
\end{align}
\end{subequations}
Note that $[e]^{}_{0i}$ and $[e]^{}_{ij}$ are projected variations with respect to $e^{}_{\mu\nu} = e^{}_\mu{}^i \, e^{}_\nu{}^i$\,.  
Moreover, in the M0T case where the strict $\omega \rightarrow \infty$ limit is considered, the associated supergravity action is a quasi one, which does not capture the extra equation of motion that arises from varying with respect to $\omega$ in the full IIA supergravity action followed by the $\omega \rightarrow \infty$ limit. Such a limit of the variation gives rise to the finite expression,
\be
    [\omega] = e^{-2 \varphi} \, \bigl( 2 \, r_{00} + 2 \, k_{0i} \, t_{0i} \bigr) + e^{-\varphi} \, \bigl( 2 \, \tau^\mu \, \nabla_{\!i} f_{\mu i} - 8 \, t_{0i} \, f_{0i} \bigr) + \tfrac{1}{2} \, f_{ij} \, f_{ij} - \tfrac{1}{6}\, \tilde{f}_{0ijk} \, \tilde{f}_{0ijk}\,. 
\ee 
See \emph{e.g.}~\cite{Bergshoeff:2024ipq} for studies of the boost transformations of the related equations of motions. When various source terms are included, the related equations of motion can be readily written down, using which one could solve for the non-Lorentzian curved geometries discussed in this paper. 

\section{Perspective from Non-Relativistic String Theory} \label{app:pnrst}

In this appendix we discuss non-relativistic string theory~\cite{Klebanov:2000pp, Gomis:2000bd, Danielsson:2000gi}, which is the first-quantized version~\cite{Blair:2024aqz, Harmark:2025ikv} of matrix string theory~\cite{Motl:1997th, Dijkgraaf:1997vv} that is dual to the BFSS matrix theory on a circle. 
Our starting point is the M0T limiting prescription~\eqref{eq:m0tcb}, which we T-dualize in a spatial isometry direction and then S-dualize to obtain~\cite{Bergshoeff:2019pij, Ebert:2021mfu, Blair:2023noj},
\begin{subequations} \label{eq:nrstlp}
\begin{align}
    G^{}_{\mu\nu} &= \omega^2 \, \tau^{}_{\mu\nu} +e^{}_\mu{}^i \, e^{}_\nu{}^i, 
        &%
    B^{(2)} &= - \omega^2 \, \tau^0 \wedge \tau^1 + b^{(2)}, \\[4pt]
    e^\Phi &= \omega \, e^\varphi, 
        &%
    C^{(q)} &= \frac{\omega^2}{e^\varphi} \, \tau^0 \wedge \tau^1 \wedge c^{(q-2)} + c^{(q)},
\end{align}
\end{subequations}
with $\tau^{}_{\mu\nu} = \tau^{}_\mu{}^a \, \tau^{}_\nu{}^b \, \eta^{}_{ab}$\,, $a = 0\,,\,1$ and $e^{}_{\mu\nu} = e^{}_\mu{}^i \, e^{}_\nu{}^i$\,, $i = 2\,,\,\cdots,\,9$\,. The $\omega \rightarrow \infty$ limit of type IIB superstring theory with the background fields parametrized as in Eq.~\eqref{eq:nrstlp} defines the IIB version of non-relativistic string theory. This is the BPS decoupling limit that zooms in on a background F-string. The fundamental degrees of freedom that are the D-particles in M0T are mapped to the F-strings in non-relativistic string theory, which is a perturbative string theory. Conveniently, the string sigma models now acquire a Lorentzian worldsheet to which standard conformal theoretic techniques are applicable~\cite{Gomis:2000bd}. Focusing on the bosonic sector, the reparametrization in $\omega$ of the Nambu-Goto action gives
\be \label{eq:slambda}
    S^{}_\lambda = - T \int \dd^2 \sigma \, \frac{\tau}{\lambda} \! \ls \sqrt{\det \lr \delta^\alpha_\beta + \lambda \, \tau^{\alpha\gamma} \, \p^{}_\gamma X^\mu \, \p^{}_\beta X^\nu \, e^{}_{\mu\nu} \rr} - 1 \rs - T \int b^{(2)},
        \quad%
    \lambda = \omega^{-2},
\ee
with $\tau^{}_{\alpha}{}^a \equiv \p^{}_\alpha X^\mu_{} \, \tau^{}_\mu{}^a$, $\tau^{}_{\alpha\beta} = \tau^{}_\alpha{}^a \, \tau^{}_\beta{}^b \, \eta^{}_{ab}$\,, and $\tau \equiv \det ( \tau_\alpha{}^a )$\,. Taking the $\lambda \rightarrow 0$ (\emph{i.e.}~$\omega \rightarrow \infty$) limit, we obtain the worldsheet action describing non-relativistic string theory~\cite{Andringa:2012uz},
\be
    S^\text{\scalebox{0.8}{NG}}_\text{nrst} = - \frac{T}{2} \int \dd^2 \sigma \, \tau \, \tau^{\alpha\beta} \p^{}_\alpha X^\mu \p^{}_\beta X^\nu \, e^{}_{\mu\nu} - T \int b^{(2)}.
\ee
At the lowest order of $\lambda$\,, the deformation from $S^{}_\text{nrst}$ to $S^{}_\lambda$ is generated by the flow equation~\cite{Zamolodchikov:2004ce, Smirnov:2016lqw},
\be
    \frac{\dd S_\lambda}{\dd \lambda} = - \frac{1}{2} \, \tau \det \bigl( T^{}_{\alpha\beta} \bigr)\,,
\ee
where $T_{\alpha\beta}$ is the stress-energy tensor associated with Eq.~\eqref{eq:slambda}. This is the standard $T\bar{T}$ flow, which in this string theory context deforms type IIB non-relativistic string theory to the full type IIB superstring theory~\cite{Blair:2020ops, Blair:2024aqz}. 

In the Polyakov action, the sigma models describing non-relativistic string theory are given by~\cite{Bergshoeff:2018yvt}
\begin{align} \label{eq:wsp}
\begin{split}
    S^\text{\scalebox{0.8}{P}}_\text{nrst} &= - \frac{T}{2} \! \int \! \dd^2 \sigma \, \sqrt{-\gamma} \, \Bigl[ \gamma^{\alpha\beta} \, \p^{}_\alpha X^\mu \, \p^{}_\beta X^\nu \, e^{}_{\mu\nu} + \epsilon^{\alpha\beta} \bigl( \chi \, \gamma^{}_\alpha \, \p^{}_\beta X^\mu \, \tau^{}_\mu + \bar{\chi} \, \epsilon^{\alpha\beta} \, \bar{\gamma}^{}_\alpha \, \p^{}_\beta X^\mu \, \bar{\tau}^{}_\mu \bigr) \Bigr] \\[4pt]
    & \quad - T \! \int \!\dd^2 \sigma \, \epsilon^{\alpha\beta} \, \p^{}_\alpha X^\mu \, \p^{}_\beta X^\nu \, b^{}_{\mu\nu} - \frac{T}{2} \! \int \! \dd^2 \sigma \sqrt{-\gamma} \, R(\gamma) \, \Phi\,.
\end{split}
\end{align}
Here, $\gamma^{}_{\alpha\beta} = - \gamma^{}_\alpha{}^0 \, \gamma^{}_\beta{}^0 + \gamma^{}_\alpha{}^1 \, \gamma^{}_\beta{}^1$, $\gamma_\alpha = \gamma_\alpha{}^0 + \gamma_\alpha{}^1$, $\bar{\gamma}_\alpha = \gamma_\alpha{}^0 - \gamma_\alpha{}^1$, $\gamma = \det \gamma^{}_{\alpha\beta}$\,, and $\gamma^{\alpha\beta}$ is the inverse of $\gamma_{\alpha\beta}$\,. Note that $\chi$ and $\bar{\chi}$ are the Lagrange multipliers imposing the constraints $\epsilon^{\alpha\beta} \, \gamma^{}_\alpha \, \p^{}_\beta X^\mu \, \tau^{}_\mu = \epsilon^{\alpha\beta} \, \bar{\gamma}^{}_\alpha \, \p^{}_\beta X^\mu \, \bar{\tau}^{}_\mu = 0$\,, with $\tau_\mu = \tau_\mu{}^0 + \tau_\mu{}^1$ and $\bar{\tau}_\mu = \tau_\mu{}^0 - \tau_\mu{}^1$\,. The $\omega$-deformation parametrized as in Eq.~\eqref{eq:nrstlp} corresponds to the current-current deformation,
\be
    - \frac{T}{2} \int \dd^2 \sigma \, \sqrt{-\gamma} \, \chi \, \bar{\chi} \, \lambda\,.
\ee
In this Polyakov formulation, the $T\bar{T}$ deformation is a marginal deformation that derives non-relativistic string theory towards the full relativistic string theory. The coupling $\lambda = \omega^{-2}$ should be promoted to a background field dependent on the embedding coordinates. The one-loop beta-function of $\lambda$ is~\cite{Gomis:2019zyu, Yan:2019xsf, Gallegos:2019icg, Yan:2021lbe} 
\be \label{eq:betalambda}
    \beta (\lambda) \, \propto \, e^{\mu\rho} \, e^{\nu\sigma} \, \p^{}_{[\mu} \tau^{}_{\nu]} \, \p^{}_{[\rho} \bar{\tau}^{}_{\sigma]}\,,
\ee
whose associated target space equation of motion is the one that is sourced in the presence of a static string. In the so-called torsional string Newton-Cartan gravity whose dynamics is determined by the vanishing Weyl anomalies of the worldsheet theory~\eqref{eq:wsp}, the beta-function of $\lambda$ given in Eq.~\eqref{eq:betalambda} corresponds to varying the target space low-energy supergravity action with respect to the Kalb-Ramond field $b_{\mu\nu}$\,. Coupled to static strings, we are led to the non-relativistic string analog of the M0T supergravity equation of motion~\eqref{eq:teom}, 
\be \label{eq:ssourcing}
    e^{\mu\rho} \, e^{\nu\sigma} \, \p^{}_{[\mu} \tau^{}_{\nu]} \, \p^{}_{[\rho} \bar{\tau}^{}_{\sigma]} \,\propto\, \frac{N \, e^{2\varphi} \, \delta^{(8)}\bigl( x^i \bigr)}{\det(\tau^{}_\mu{}^0 \,\,\, \tau_\mu{}^1 \,\,\, e^{}_\mu{}^i)}\,.
\ee
However, it is evident from the worldsheet theory that the inclusion of a non-trivial torsion such that the beta-function~$\beta(\lambda)$ does not vanish (at the source) already generates a renormalization group flow towards the full string theory, which implies that the background field $\lambda$ should have been included from the very beginning. This leads us back to the standard IIB supergravity as the low-energy effective theory, whose equations of motion with a now relativistic string source are well-known to be solved by $\lambda = \omega^{-2} \sim (\ell / r)^{6}$, with all the other background fields corresponding to the flat solution in torsional string Newton-Cartan gravity~\cite{Danielsson:2000mu, Avila:2023aey, Guijosa:2023qym, Blair:2024aqz, Harmark:2025ikv}. This behavior is analogous to the dual M0T case discussed at the beginning of this section, and may be viewed as a further corroboration using the string worldsheet. Moreover, this also makes it necessary to think about an expansion instead of a limit of string theory as in~\cite{Hartong:2021ekg, Hartong:2022dsx, Hartong:2024ydv}.

The above discussion raises the question: in what sense can one define a self-contained non-relativistic string theory? We return to the construction of the sigma model~\eqref{eq:wsp} for non-relativistic string theory, but now construct it from symmetry principles. Focusing on the bosonic sector, we demand that the theory is invariant under the background field transformations generated by the extended string Galilei algebra consisting of longitudinal translation $H_a$\,, transverse translations $P_{i}$\,, longitudinal Lorentz boost $M$\,, transverse rotations $J_{ij}$\,, string Galilei boosts $G_{ai}$\,, and the extensions including a string winding charge $Z_a$ analogous to the particle number generator in the Bargmann algebra. Apart from the standard commutators in a stringy generalization of the Galilei algebra, there is also a non-trivial commutator between the transverse translations and string Galilei boosts, $[P_i\,, G_{aj}] = \delta_{ij} \, Z_a$~\cite{Brugues:2006yd, Andringa:2012uz, Bergshoeff:2019pij}.\,\footnote{Allegedly, this symmetry algebra arises from the BPS decoupling limit adapted towards non-relativistic string theory~\cite{Bidussi:2021ujm, Blair:2024aqz}.} Under the action of $Z_a$\,, we have $\delta b_{\mu\nu} = \tau_{[\mu}{}^a \, D_{\nu]} \sigma^b \, \epsilon^{}_{ab}$\,, with $\sigma^a$ the Lie group parameter associated with the generator $Z_a$ and $D_\mu$ the covariant derivative with respect to the frame index $a$\,. Demanding the invariance of the string sigma model~\eqref{eq:wsp} under the $\sigma^a$ transformation imposes the torsional constraint $D_{[\mu} \tau_{\nu]}{}^a = 0$\,. Under this condition, the beta-function of $\lambda$ vanishes perturbatively at all loop orders. 
Nevertheless, backreaction the F-string will explicitly break the $Z_a$ symmetry and turn on the $T\bar{T}$ deformation that derives non-relativistic string theory to type IIB superstring theory. 

Note that the finiteness of the supersymmetry transformation in the non-relativistic string limit prescribed by Eq.~\eqref{eq:nrstlp} requires certain torsional constraints. In the $\CN = 1$ case (\emph{i.e.}~heterotic supergravity) half of the torsional constraint~$D_{[\mu} \tau_{\nu]}{}^a = 0$ is required~\cite{Yan:2021lbe, Bergshoeff:2021tfn, Bergshoeff:2023fcf, Bergshoeff:2025uut},\,\footnote{This chiral version of the geometric constraint is also required for the quantum consistency associated with the gravitational anomaly in heterotic non-relativistic string theory~\cite{hsa}.} and it is suspected that in the $\CN = 2$ case (\emph{i.e.} IIA/B supergravity) the full torsional constraint~$D_{[\mu} \tau_{\nu]}{}^a = 0$ might be required~\cite{Bergshoeff:2024nin}. This suggests that the analogous torsional constraint $\p_{[\mu} \tau_{\nu]}$ in the M0T supergravity may have to hold in order to preserve the supersymmetry, which deserves further analysis of the full non-Lorentzian supergravity, with the fermionic sector included. 


\bibliographystyle{JHEP}
\bibliography{nsmt}

\end{document}